\documentclass[prx,reprint,superscriptaddress,twocolumn,showkeys]{revtex4-1}
\usepackage[utf8]{inputenc}
\usepackage{amsmath}
\usepackage{braket}
\usepackage{graphicx}
\usepackage{pgfplots}
\pgfplotsset{compat=1.15}
\usepackage{csquotes}
\usepackage{hhline}
\usepackage{amssymb}

\usepackage{dcolumn}
\usepackage{tabularx}
\usepackage[colorlinks=true,linkcolor=blue,citecolor=blue,urlcolor=blue]{hyperref}
\usepackage{braket}
\usepackage{float}
\usepackage{listings}
\usepackage{color}
\usepackage{subcaption}
\usepackage{placeins}
\definecolor{mygreen}{rgb}{0,0.6,0}
\definecolor{mygray}{rgb}{0.5,0.5,0.5}
\definecolor{mymauve}{rgb}{0.58,0,0.82}

\newcolumntype{C}{>{\centering\arraybackslash}X}

\begin{document}

\title{Demonstration of Optimal Fixed-Point Quantum Search Algorithm in IBM Quantum Computer}

\author{Bikramaditya Das}
\email{dasbikramaditya2@gmail.com}
\affiliation{Department of Chemistry, \\University of Calcutta, 92 APC Road, Kolkata 700009, West Bengal, India}
\author{Kamal Gurnani}
\email{kamalgurnani@iitkgp.ac.in}
\affiliation{Department of Physics, \\Indian Institute of Technology Kharagpur, Kharagpur 721302, West Bengal, India}
\author{Bikash K. Behera}
\email{bikas.riki@gmail.com}
\affiliation{Department of Physical Sciences,\\ Indian Institute of Science Education and Research Kolkata, Mohanpur 741246, West Bengal, India}
\author{Prasanta K. Panigrahi}
\email{pprasanta@iiserkol.ac.in}
\affiliation{Department of Physical Sciences,\\ Indian Institute of Science Education and Research Kolkata, Mohanpur 741246, West Bengal, India}

\begin{abstract}
\textbf{Abstract:} A quantum algorithm is a set of instructions for a quantum computer, however, unlike algorithms in classical computer science their results cannot be guaranteed. Quantum search algorithm can be described as the rotation of state vectors in a Hilbert space. The state vectors uniformly rotate by iterative sequences until they hit the target position. To optimize the algorithm, it is necessary to have the precise knowledge about some parameters like the number of target positions and total number of states. Here, we demonstrate the implementation of optimal fixed-point quantum search algorithm in IBMQ simulator developed by IBM corporation. We perform the search algorithm for one-iteration for two, three, four and five qubit systems and confirm the accuracy of our results through histogram.
\end{abstract}


\maketitle

\section{Introduction}
Quantum versions of classical algorithms have always been a boon for the typical problems which are encountered in various branches of computation and information processing \cite{nielsen2002quantum, bennett2000quantum} i.e., data security \cite{childs2001secure, diamanti2017best}, cryptography \cite{ekert1991quantum, gisin2002quantum, bennett1992quantum}, database handling \cite{maitra2017data}, to name a few. Many of these exploit the simple but crucial fact that in quantum theory, a superposition of states can also exist at a time unlike classical domain where only a single state (string of bits) can be represented at an instance \cite{brickman2005implementation}. Almost all quantum algorithms can be expressed in a query (oracle) model where the input is given by a black box which answers queries of a certain form. In the query model, the input \(x_1,x_2,...,x_n\) is contained in a black box and can be accessed by queries to the black box. In each query, we give $i$ to the black box, and the black box outputs \(x_i\). The goal is to solve the problem with the minimum number of queries. In the first case it has two inputs $i$ and $j$, where $i$ consisting \(\log_2 N\) bits and $j$ consisting of 1 bit. If the input to the query box is the basis state \(\ket{i}\ket{j}\), the output is \(\ket{i}\ket{j\bigoplus x_i}\), where \(\bigoplus\) denotes addition modulo 2. If the input is superposition of states \(\sum_{i,j} a_{i,j}\ket{i}\ket{j\bigoplus x_i}\). In the second of quantum query,the black box has just one query $i$. If the input is a state \(\sum_i a_i\ket{i}\),then the output will be \(\sum_i a_i (-1)^{x_i}\ket{i}\). This query model, for example, serves as the basis of Grover's search algorithm \cite{pan2019operator}.

Grover's search algorithm \cite{chuang1998experimental} for unsorted database is another brilliant example of that. This algorithm comprising of the iterative applications of the Grover's operator \cite{pan2019operator}, has a beautiful geometrical understanding when the initial state being a superposition of all the states present in the database is expressed in a 2-dimensional Hilbert space with the orthogonal states being (a) superposition of unmarked or non-target states (b) superposition of marked or target sates. Thus, Grover's operator \cite{pan2019operator} is a unitary operator which serves to rotate the given initial state towards the target state axis in this Hilbert space effectively increasing the amplitude of the target state in the initial superposition \cite{friedman2000quantum}. This algorithm when applied within an unsorted database of N items having M marked items, can perform the task in \(\sqrt{N/M}\) applications of the Grover's operator, decomposition of which depends upon the requirement of search problem. However, its usefulness is limited as a general problem does not come up with the known number of marked items without which the number of iteration is unknown to us. This leads to us the souffle problem i.e. only a few iterations leave the state mostly comprising of the unmarked states while too many iterations cause the state vector to surpass the target state in the assumed 2-D Hilbert space.

Souffle problem \cite{behrens2011perfect} is dealt with using the fixed-point quantum search (FPQS) algorithms which always increases the amplitude of the target state with each iteration. One such (FPQS) algorithm is Grover's \(\frac{\pi}{3}\) algorithm. But these algorithms lose the quadratic speed-up which is the astonishing and useful feature of quantum algorithms. Furthermore, Yoder \emph{et al.} \cite{yoder2014fixed} developed another algorithm which avoids the souffle problem with the quadratic speed up. This requires setting the success probability with a bound over it in the form of a tunable parameter \( \delta\). In this case, we can achieve the quadratic speed-up as well in our process. This algorithm, abbreviated as OFPQS (Optimal FPQS) \cite{bhole2016steering} has been discussed in  section \ref{SecII}.

IBM Quantum Experience is widely used to perform different tasks in the field of quantum computation and quantum information \cite{rundle2017simple, grimaldi2001distributed, kalra2019demonstration, ghosh2018automated, gangopadhyay2017generalization, li2017approximate, sisodia2017design, joy2017experimental, yalccinkaya2017optimization, dash2017quantum, roy2017experimental, satyajit2018nondestructive, li2017approximate, huffman2017violation, alsina2016experimental, wootton2017demonstrating, berta2016entropic, deffner2017demonstration, garcia2017five}. Experimental test of Hardy’s paradox \cite{das2020new}, topological quantum walks \cite{balu2018physical}, quantum permutation algorithm \cite{yalccinkaya2017optimization} have been illustrated. Error correction with 15 qubit repetition code \cite{wootton2017demonstrating} and estimation of molecular ground state energy have also been implemented using 16 qubit IBM quantum computer, ibmqx5. Alvarez-Rodrigue \emph{et al}. \cite{alvarez2016artificial} have shown artificial life in quantum technologies. Current trends like quantum machine learning \cite{schuld2017implementing} has been performed on the quantum computer. An essential ingredient of quantum communication, quantum repeater has been designed by Behera \emph{et al}. \cite{behera2019demonstration} using IBM quantum computer. One of the important quantum mechanical problems, quantum tunneling \cite{hegade2017experimental} has been simulated on the universal quantum simulator, ibmqx4 and IBM Q 14 Melbourne. Yoder \emph{et al.} proposed an algorithm also \cite{yoder2014fixed, bhole2016steering} with the oracle designed for the desired state. In this paper, we demonstrate the OFPQS algorithm in IBMQ QASM simulator by taking two, three, four and five qubits systems and taking one target state at a time.

The rest of the paper is organized as follows. In the section \ref{SecII}, we explain the OFPQS and Grover's search algorithm. In section \ref{SecIII}, we explain the experimental procedures of OFPQS algorithm and we demonstrate the circuit and histograms ( i.e., probabilities of final states) of the states of two, three, four and five qubit systems.

\section{OFPQS And Grover's Search Algorithm \label{SecII}}
\begin{figure}
\centering
\includegraphics[scale=0.4]{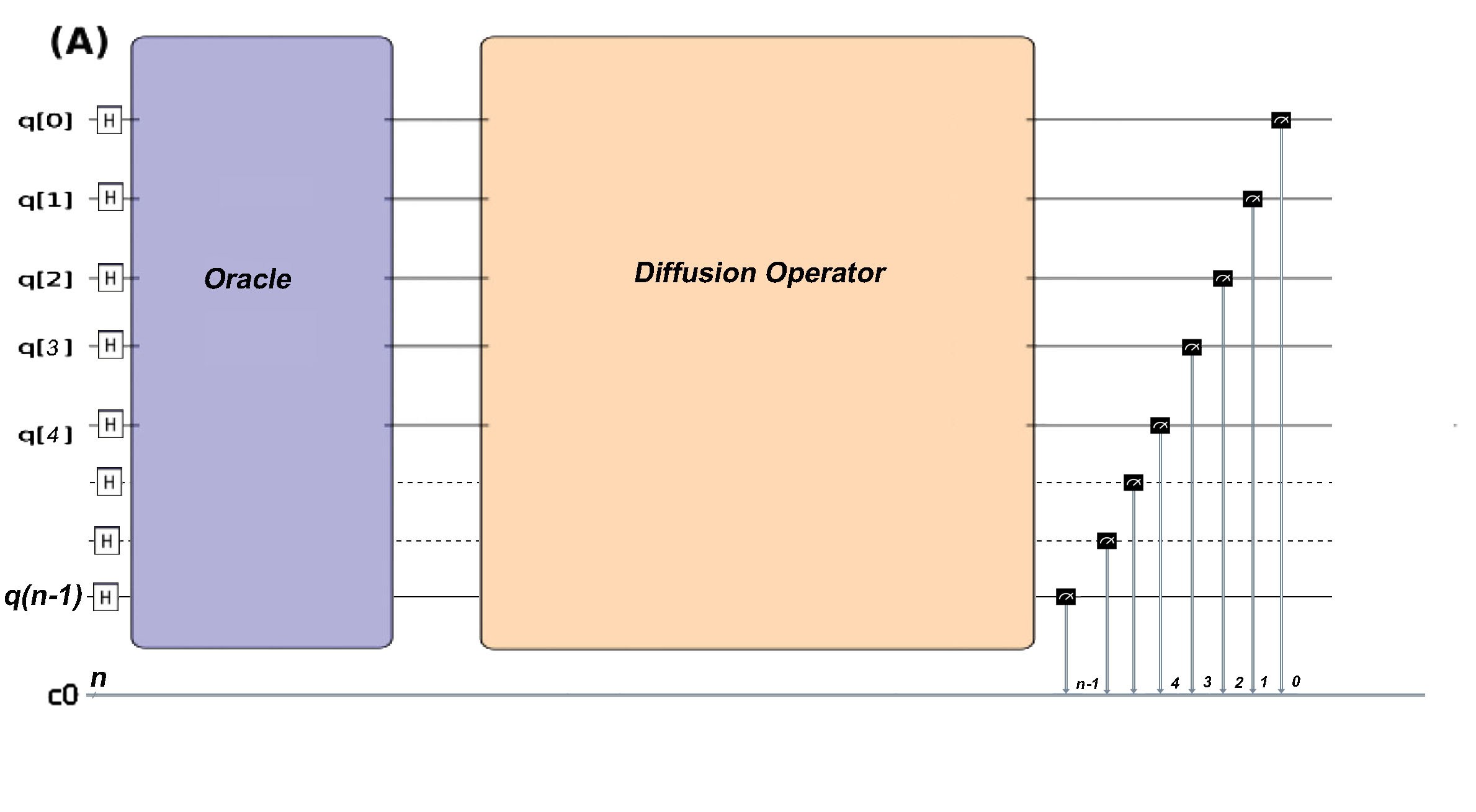}
\caption{ Generalised quantum circuit for optimal fixed-point quantum search algorithm for n-qubit system: Here, initially the oracle is implemented on the equal superposition state of all possible n-qubit states and after that, the diffusion operator is implemented, finally all qubits are measured. }
\label{Schematic.jpg}
\end{figure}

Classical search algorithms can find one or more ‘marked’ items among an unsorted database of N items in O(N) steps. On the other hand, Grover’s quantum search algorithm \cite{pan2019operator} achieves the same task in $ O(\sqrt{N})$ steps, thereby providing a quadratic speedup over the classical counterpart \cite{sisodia2017design}. Grover’s algorithm identifies one of the M marked items among N unsorted items with the help of a given oracle function that can recognize the marked items. It can also be interpreted as a rotation in the 2D space spanned by the superposition of non-solution states and the superposition of M solution states. The application of the Grover iteration can thus be visualized as the rotation of the initial state towards the target state in $ O(\sqrt{N/M}) $ steps. However,if we do not know the number of marked items M beforehand, we cannot predict the number of iterations which would land the initial state closest to the marked state \cite{pan2019operator}. Too few iterations give us a state comprising of mostly non-solution states, whereas, too many iterations can surpass the solution states and we may end up getting non-solution states, yet again. In order to overcome this problem, attempts have been made to develop fixed point quantum search (FPQS) algorithms, which monotonically amplify the probability of obtaining the marked states \cite{yalccinkaya2017optimization, dash2017quantum}. The Grover's search algorithm solves the problem. The Grover's search algorithm is optimal and it is an optimal fixed-point quantum search algorithm. Here in this article, we show the demonstration of the optimal fixed-point quantum search algorithm for 2, 3, 4, 5-qubits systems for better understanding of the application of this algorithm. In this article, we outline the various steps involved in the OFPQS algorithm \cite{pan2019operator}. A generalised quantum circuit for n-qubit system is shown in Fig. \ref{Schematic.jpg}. The oracle and diffusion operator combine to form the Grover's operator for single iteration \cite{pan2019operator}. In Grover's search algorithm, the Grover's operator is implemented $\frac{\pi}{4} \sqrt{N}$ times to reach the final state with heighest probability. Here in OFPQS algorithm, we implement the oracle and then the diffusion operator once, hence only we perform single iteration to reach the final target state. In this process, we choose proper two and multi-qubit gates to reach the final states from the equal superposition state in a single iteration only.

\textbf{Search Problem for an Unstructured Database:} 
Find the record \(\omega\) with the minimum amount of computational work, i.e., with the minimum number of queries. To solve the Grover's search problem suppose we are supplied with a quantum oracle with the ability to recognize solution to the search problem. The oracle is a unitary operator U, defined by its action on the computational basis:

\begin{eqnarray} 
U: \ket{x}\ket{q} \rightarrow \ket{x}\ket{q \oplus f(x)}
\end{eqnarray} 

where \(\ket{x}\) is the input register, \(\ket{q}\) is the oracle qubit, and \(\oplus\) denotes addition modulo 2. The oracle qubit \(\ket{q}\) is a single qubit which is flipped if f(x) = 1, and is unchanged otherwise. We can check whether x is a solution to our search problem by preparing \(\ket{x}\ket{0}\), applying the oracle and checking to see if the oracle qubit has been flipped to \(\ket{1}\). It is useful to choose the state of the single - qubit register to be: \(\frac{1}{\sqrt 2} (\ket{0} - \ket{1})\). We achieve this state simply applying the Hadamard transform to the \(\ket{1}\) quantum state:

\begin{eqnarray} 
H &=& \frac{1}{\sqrt 2}
\begin{bmatrix}
1 & 1\\
1 & -1
\end{bmatrix}\nonumber\\
\ket{0} &\xrightarrow{H}& \frac{1}{\sqrt 2} (\ket{0} + \ket {1})\nonumber\\
\ket{1} &\xrightarrow{H}& \frac{1}{\sqrt 2} (\ket{0} - \ket {1})
\end{eqnarray}

Now we apply the oracle to the new state:
\begin{eqnarray} 
U\ket{x}\frac{(\ket{0}- \ket{1})}{\sqrt 2} = (-1)^{f(x)} \ket{x}\frac{(\ket{0}- \ket{1})}{\sqrt 2}
\end{eqnarray}

For instance, if $x$ is a solution to the search problem then the final state will be: \begin{eqnarray}
-\ket{x}\frac{(\ket{0}- \ket{1})}{\sqrt 2}
\end{eqnarray}. It is obvious, that the state of the oracle qubit is not changed, so we may ignore it, and obtain: 

\begin{eqnarray}
\ket{x}\xrightarrow{U} (-1)^{f(x)} \ket{x}
\end{eqnarray}

It is easy to see, that the oracle transformation U is equivalent with the following transformation: \(U= I-2\ket{\omega}\bra{\omega}\) is the only solution of the search problem, and I is the identity matrix. If \(\ket{x}=\ket{\omega}\), then \(U\ket{\omega}= I\ket{\omega} -2\ket{\omega}\bra{\omega}\ket{\omega} = I\ket{\omega} - 2\ket{\omega}= - \ket{\omega}\). If \(\ket{x}\neq \ket{\omega}\), then \(U\ket{x}= I\ket{x} -2\ket{\omega}\bra{\omega}\ket{x} = I\ket{x}\), since \(\ket{\omega}\) and \(\ket{x}\) are orthogonal to each other if \(\ket{\omega} \neq \ket{x}\). Thus the oracle U flips the sign of \(\ket{x}\) if \(\ket{x}=\ket{\omega}\) but operates trivially on all \(\ket{x}\neq \ket{\omega}\). More precisely, when U operates on some vector \(\ket{x}\) in Hilbert space \(H^{2^n} (N=2^n)\), reflects it about the hyperplane orthogonal to \(\ket{\omega}\). We know that the reflection is performed for some basis state \(\ket{\omega}\) , but nothing is known about the string \(\omega\) itself. Our task is to determine \(\omega\). To achieve our aim, we are using the so called Grover iteration: \cite{pan2019operator, borbely2007grover}.

\begin{itemize}
\item Create a perfect random state \(\ket{\Psi} = \frac{1}{\sqrt {2^n}} \sum_{i=0}^{2^n -1} \ket{x_i}\) by application of the Hadamard transformation on the state \(\ket{0}^{\otimes n}\). We know about \(\ket{\omega}\) that it is a base state, which means that \(\ket{\Psi}\bra{\omega}\) Measuring  the state \(\ket{\Psi}\) by projection onto its base {\(\ket{x}\)} would return the state \(\ket{\omega}\) only with  probability \(\frac{1}{N}\).

\item Combine the unknown reflection U with some known reflection \( V = 2\ket{\Psi}\bra{\Psi} - I\). It 
flips the sign of \(\ket{x}\) if it is orthogonal to \(\ket{\Psi}\) and preserves the sign of \(\ket{x}\) , if  \(\ket{x}=\ket{\Psi}\): \( V\ket{x} = 2\ket{\Psi}\bra{\Psi}\ket{x} - I\ket{x} = - I\ket{x}\), since \(\ket{\Psi}\) and \(\ket{x}\) are orthogonal to each other and eventually \(\bra{\Psi}\ket{x}=0\). \( V\ket{\Psi} = 2\ket{\Psi}\bra{\Psi}\ket{\Psi} - I\ket{\Psi} = 2\ket{\Psi} - I\ket{\Psi} = \ket{\Psi}\), since \(\ket{\Psi}\) is normalised and eventually \(\bra{\Psi}\ket{\Psi}=1\). Acting on some arbitrary vector, V preserves the component along \(\ket{\Psi}\) but flips all components in the hyperplane orthogonal to \(\ket{\Psi}\).

\item Apply the operator \(G=V*U\) to some vector \(\ket{x}\) with the following effect: \cite{pan2019operator}\\ 
    \begin{itemize}
        \item the vector \(\ket{\Psi}\)  and \(\ket{\omega}\) span a plane in \(H^{\otimes N}\) with a vector \(\Ket{\omega ^ \perp}\) that is orthogonal to \(\ket{\omega}\) in that plane;
        \item any vectors \(\ket{x}\) in that plane is by U flipped about \(\ket{\omega ^ \perp}\) and by V flipped about \(\ket{\Psi}\).
\end{itemize}

To understand what this means, consider the geometry in a two-dimensional Hilbert space: (Fig. \ref{qdctc_Fig4}) \cite{borbely2007grover}.

\begin{itemize}
\item The angle between \(\ket{\Psi}\) and \(\ket{\omega}\) is given as:
\begin{eqnarray}
\bra{\Psi}\ket{\omega}=\frac{1}{\sqrt N}=\frac{1}{\sqrt {2^n}}=\cos{(\frac{\pi}{2}-\theta})\
\end{eqnarray}
\item A vector \(\ket{x}\) is by U flipped about \(\ket{\omega^\perp}\) and thereby changed its angle by 2\(\phi\)
\item The vector ' \(U\ket{x}= \ket{x}\) is by V flipped about \(\ket{\Psi}\), thus changing its angle by 2\(\delta\)
\item The total change of angle effected by G=V*U is thus \(2\phi + 2\delta = 2\theta\), which is the effect of one Grover iteration.
\end{itemize}

\begin{figure*}[!ht]
\centering
\begin{subfigure}{0.3\linewidth}
\includegraphics[width=\linewidth]{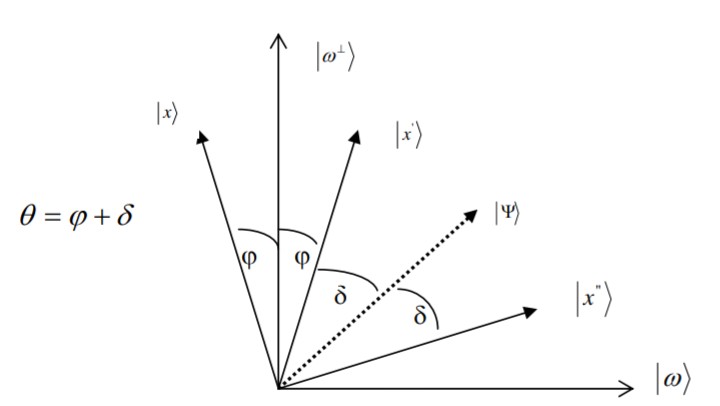} 
\caption{}
\label{qdctc_Fig1}
\end{subfigure}\hfill
\begin{subfigure}{0.3\linewidth}
\includegraphics[width=\linewidth]{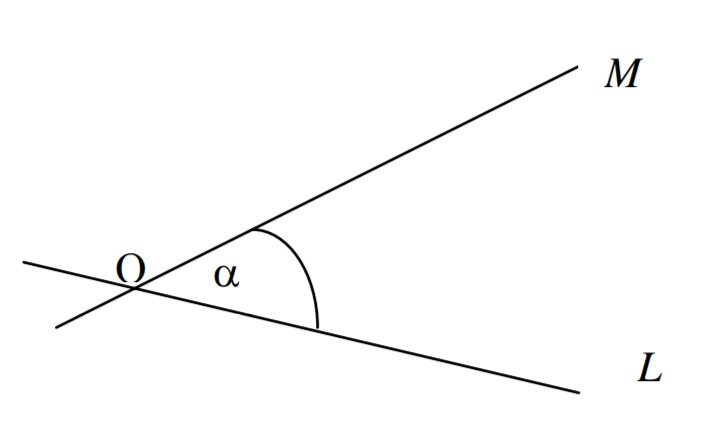} 
\caption{}
\label{qdctc_Fig2}
\end{subfigure}\hfill
\begin{subfigure}{0.3\linewidth}
\includegraphics[width=\linewidth]{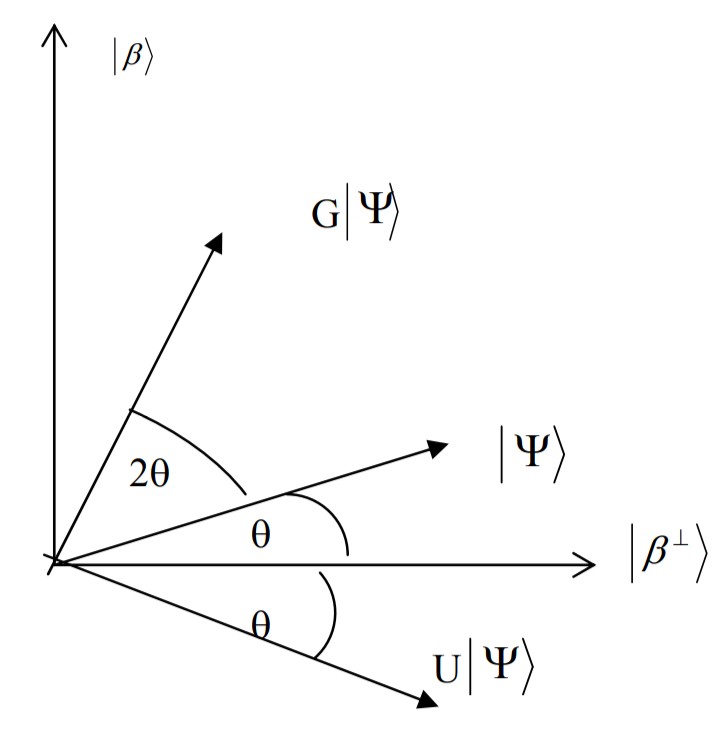} 
\caption{}
\label{qdctc_Fig3}
\end{subfigure}\hfill
\caption{{(a) Flipping of a vector upon application of G operator. (b) This figure suggests the operation of reflection in L followed
by reflection in M is just a rotation by angle \(2\alpha\) about the point O. (c) This figure demonstrates \(\ket{\Psi}\), \( G \ket{\Psi}\), \( U \ket{\Psi}\) properly.}}
\label{qdctc_Fig4}
\end{figure*}

\textbf{Remark:}
The Grover iteration \cite{byrnes2018generalized} may be seen to be a consequence of the following elementary theorem of 
2-dimensional real Euclidian geometry.

\textbf{\underline{Theorem}:} Let L and M be two mirror lines in the Euclidian plane \textbf{\(R^2\)} intersecting at a point O and let \(\alpha\) be the angle in the plane from L to M. Then the operation of reflection in L followed by reflection in M is just a rotation by angle \(2\alpha\) about the point O (Fig. \ref{qdctc_Fig4}) \cite{borbely2007grover}.

\item Repeat the Grover iteration until the possibility the input item in the data base 
approaches the value 1. We assume that we need some k iteration to bring \(\ket{\Psi}\) close to \(\ket{\omega}\) (or to turn it away from \(\ket{\omega^\perp}\) by an angle of \(\frac{\pi}{2}\). After k iteration we thus get a rotation by \(\theta + 2k\theta = \frac{\pi}{2}\). Since \(\sin{\theta} \cong \theta = \frac{1}{\sqrt N}\), for large N, we iterate until:

\begin{eqnarray}
(2k+1)\theta \cong \frac{\pi}{2} \Rightarrow k &=& round (\frac{\pi}{4\theta} - \frac{1}{2})\nonumber\\
k&=& \frac{\pi}{4} \sqrt{N}
\end{eqnarray}

Only \(k= \frac{\pi}{4} \sqrt{N}\) no of iterations are required to find \(\ket{\omega}\) with high probability. For example, assuming that the data base contains just 4 items. A classical sequential search would requires 2 steps to find a particular item. A quantum-mechanical search for an item \(\ket{x}\) would set out \(|\bra{x}\ket{\omega}| = \frac{1}{\sqrt 4} = \frac{1}{2} = \sin{\theta}\) which means that \(\theta = \sin^{-1}{(\frac{1}{2})} = 30 ^\circ\). A rotation by \(2\theta = 60^\circ\) brings  \(\ket{x}\) into a \(90^\circ\) angle with \(\ket{\omega^\perp}\), or in line with \(\ket{\omega}\).

\end{itemize}

\begin{itemize}
\item \textbf{Multiple Solutions:}\\
Suppose that the search problem has exactly M solutions, with \(1\leq M \leq N\),\(N= 2^n\) and M is known. In this case the oracle introduces a reflection in the hyperplane orthogonal to the vector \(\ket{\beta}=\frac{1}{\sqrt M}\sum_{i=1}^M \ket{\omega_i}\),  or in other manner \(\ket{\beta}=\frac{1}{\sqrt M}\sum_{x=f^{-1} (1)} \ket{x}\), the equal weighted superposition of the marked computational basis states. The original state \(\ket{\Psi}= \frac{1}{\sqrt N}\sum_{i=0}^{N-1} \ket{x}\) can be rewritten as,

\begin{eqnarray}
\ket{\Psi} &=& \frac{\sqrt {N-M}}{\sqrt N}\bigg(\frac{1}{\sqrt {N-M}}\sum_{f^{-1}(0)} \ket{x}\bigg)\nonumber\\ 
&+& \frac{\sqrt M}{\sqrt N}\bigg(\frac{1}{\sqrt M} \sum_{f^{-1}(1)}\ket{x}\bigg)\nonumber\\
&=& \frac{\sqrt {N-M}}{\sqrt N}\ket{\alpha} + \frac{\sqrt M}{\sqrt N}\ket{\beta},
\end{eqnarray}

where 

\begin{eqnarray}
\ket{\alpha} &=& \bigg(\frac{1}{\sqrt {N-M}}\sum_{f^{-1}(0)} \ket{x}\bigg)\nonumber\\
\ket{\beta} &=& \bigg(\frac{1}{\sqrt M} \sum_{f^{-1}(1)}\ket{x}\bigg)
\end{eqnarray}

Using the simplest notation, \(\ket{\Psi} = \cos\theta \ket{\alpha} + \sin\theta \ket{\beta}\) [Fig. \ref{qdctc_Fig4}]. The effect of $G = V*U$ is the following:

\begin{itemize}
\item The oracle U performs a reflection about the vector \(\ket{\alpha}\) which is orthogonal to \(\ket{\beta}\). \(U(\cos\theta \ket{\alpha} + \sin\theta \ket{\beta}) = (\cos\theta \ket{\alpha} - \sin\theta \ket{\beta})\)
\item The vector \(\cos\theta \ket{\alpha} - \sin\theta \ket{\beta}\) is by \(V = 2\ket{\Psi}\bra{\Psi} - I\) flipped about \(\ket{\Psi}\).
\item The product of two reflection is a rotation and after k iteration the state:
\begin{eqnarray}
G^k \ket{\Psi} = \cos(2k+1)\theta \ket{\alpha} + \sin(2k+1)\theta \ket{\beta}
\end{eqnarray}
\end{itemize}

For large N, when M is very much lesser than N, we have \(\theta \approx \sin\theta \approx \frac{\sqrt M}{\sqrt N} \)

\begin{eqnarray}
(2k+1)\theta&=&\frac{\pi}{2}\nonumber\\
&\Downarrow&\nonumber\\
(2k+1)\frac{\sqrt M}{\sqrt N} &=& \frac{\pi}{2}\nonumber\\
&\Downarrow&\nonumber\\
k &=& round \big( \frac{\pi}{4}\frac{\sqrt N}{\sqrt M} - \frac{1}{2}\big) 
\end{eqnarray}

So the state is close to \(\ket{\beta}\) after a number of iteration \(k \approx \frac{\pi}{4}\frac{\sqrt N}{\sqrt M}\) with probability: Probability = \(|\sin^2(2k+1)\theta|\). Reverting to the previous example; the probability to find one marked item from 4 is probability =\(|\sin^2(2k+1)\theta| = |\sin^2 3\theta| = |\sin^2 3.30^\circ| = 1\) [since k \(\approx\) 1 there] \cite{diao2002quantum, borbely2007grover}
\end{itemize}

\textbf{Quantum Search Algorithm in detail:}\\

Input:

\begin{itemize}
\item  \(x_1 = \{0,1\}, x_2 = \{0,1\}, x_3 = \{0,1\},..., x_n = \{0,1\}\); \(f: \{0,1\}^N \rightarrow \{0,1\}\); \(f(x) = 0\) \(\forall x \in [0,N]\), except \(\omega = x_i\), for which \(f(x_i) = 1\).
\item n qubits in the state \(\ket{0}\) and one qubit i.e., the oracle qubit is in the state \(\frac{(\ket{0}-\ket{1})}{\sqrt 2}\).
\end{itemize}

Procedure:
\begin{itemize}
\item The initial state is \(\ket{0}^{\otimes n}\ket{\frac{(\ket{0}-\ket{1})}{\sqrt 2}}\).\\
\item Apply the \(H^{\otimes n} \) for the first n qubits \(\Rightarrow \frac{1}{\sqrt N}\sum_0^{N-1} \ket{x} \big[\frac{(\ket{0}- \ket{1})}{\sqrt 2}\big]\)\\
\item Apply the Grover's iteration \( k \approx \frac{\pi}{4} \sqrt{N}\) times.\\
\(\Rightarrow\) \([(2\ket{\Psi}\bra{\Psi}-I)(I-2\ket{\omega}\bra{\omega})]^k \frac{1}{\sqrt N}\sum_0^{N-1} \ket{x} \big[\frac{(\ket{0}- \ket{1})}{\sqrt 2}\big] \approx \ket{\omega}\big[\frac{(\ket{0}- \ket{1})}{\sqrt 2}\big]\)\\
\item Measure first n qubits \(\Rightarrow \ket{\omega}\)
\end{itemize}
\begin{itemize}
\item \textbf{Example of N=4 (Matrix representation)}:\\
In this case, we need n=2 qubits and another one with 1 qubit, the oracle qubit. We need just at k = round\((\frac{\pi}{4} \sqrt{N}) = 1 \)oracle query to resolve the search problem. The initial states \(\ket{\psi_0} = \ket{00}\), and after applying the Hadamard gates we get:

\begin{eqnarray}
\ket{\Psi} &=& H^{\otimes 2}\ket{00}= (H\ket{0})^{\otimes 2} \nonumber\\
&=& \Bigg[ \frac{1}{\sqrt 2} \begin{bmatrix}
1 & 1\\
1 & -1
\end{bmatrix}
\begin{bmatrix}
1\\
0
\end{bmatrix}
\Bigg]^{\otimes 2}
= \Bigg[\frac{1}{\sqrt 2} \begin{bmatrix}
1\\
1
\end{bmatrix}
\Bigg]^{\otimes 2}\nonumber\\
&=& \frac{1}{2} \begin{bmatrix}
1\\
1
\end{bmatrix}
\bigotimes
\begin{bmatrix}
1\\
1
\end{bmatrix}
= \frac{1}{2}\begin{bmatrix}
1\\
1\\
1\\
1
\end{bmatrix}
\end{eqnarray}

Suppose that \(f(3) = 1\) and \( f(i) = 0\) for \(i \neq 3\) .\\
Then the oracle transformation is the identity matrix with the 3rd element of the diagonal equals to (–1).
\begin{eqnarray} U =\begin{bmatrix}
1 & 0 & 0 & 0\\
0 & 1 & 0 & 0\\
0 & 0 & -1 & 0\\
0 & 0 & 0 & 1
\end{bmatrix}
\
\end{eqnarray}

In the next step we apply the oracle transformation U to the state \(\ket{\Psi}\)
\begin{eqnarray}
\ket{\Psi_1} = U\ket{\Psi} = \begin{bmatrix}
1 & 0 & 0 & 0\\
0 & 1 & 0 & 0\\
0 & 0 & -1 & 0\\
0 & 0 & 0 & 1
\end{bmatrix}
\frac{1}{2}
\begin{bmatrix}
1\\
1\\
1\\
1
\end{bmatrix}
= \frac{1}{2}\begin{bmatrix}
1\\
1\\
-1\\
1
\end{bmatrix}
\end{eqnarray}

We have seen that the \(V = 2\ket{\Psi}\bra{\Psi} - I\) operator is equivalent with the so-called ``diffusion operator". D = \(\frac{2}{N} -1\), if \(i=j\). Again, D= \(\frac{2}{N}\), if \(i \neq j\). or in this case,

\begin{eqnarray}
D &=& 
\begin{bmatrix}
-\frac{1}{2} & \frac{1}{2} & \frac{1}{2} & \frac{1}{2}\\
\frac{1}{2} & -\frac{1}{2} & \frac{1}{2} & \frac{1}{2}\\
\frac{1}{2} & \frac{1}{2} & -\frac{1}{2} & \frac{1}{2}\\
\frac{1}{2} & \frac{1}{2} & \frac{1}{2} & -\frac{1}{2}
\end{bmatrix}\nonumber\\
\ket{\Psi_2} = D{\ket{\Psi_1}} &=&  
\begin{bmatrix}
-\frac{1}{2} & \frac{1}{2} & \frac{1}{2} & \frac{1}{2}\\
\frac{1}{2} & -\frac{1}{2} & \frac{1}{2} & \frac{1}{2}\\
\frac{1}{2} & \frac{1}{2} & -\frac{1}{2} & \frac{1}{2}\\
\frac{1}{2} & \frac{1}{2} & \frac{1}{2} & -\frac{1}{2}
\end{bmatrix}
\frac{1}{2}
\begin{bmatrix}
1\\
1\\
-1\\
1
\end{bmatrix}
\end{eqnarray}

Here, according to the Grover iteration, the amplitude of the marked state increased by \((\frac{1}{2})\), while the amplitude of other states decreased to zero. Now, if we measure the final state \(\ket{\Psi_2}\), then we obtain the correct answer with unit probability.
\end{itemize}

\section{Experimental Procedures and Results \label{SecIII}}
In this paper, we show how we can search a particular state containing a particular sets of informations through an algorithm that is like Grover's search algorithm (We follow the steps of Grover's search algorithm except performing it only for single iteration and using two $\&$ multi-qubit gates), which is an optimal fixed-point quantum search algorithm. We use upto 5-qubits systems to demonstrate the quantum search algorithm. We know that in the Grover's search algorithm, we initially employ one Hadamard gate each on each qubit, then once the equal superposition of all the states in that system is generated by those Hadamard transformations, we implement an oracle on that equal superposition of all the states. The given oracle query flips the marked/target state and makes its amplitude just \textbf{NEGATIVE}, then the diffusion operator `D' operates and leads to the marked state only as the final state with probability `1'. Here, we run our quantum circuits on the IBMQ QASM simulator. We elaborately demonstrate and explain the quantum circuits for all the target states, the algorithm and the results for the states of 2,3,4,5-qubits systems for single iteration here in this paper. We design the required quantum circuits and simulate those on IBMQ QASM simulator by writing python codes on Qiskit.

\subsection{Two qubits system}
\begin{figure*}[!ht]
\centering
\begin{subfigure}{0.3\linewidth}
\includegraphics[width=\linewidth]{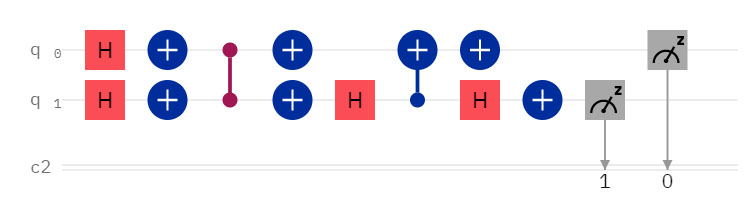} 
\caption{}
\label{qdctc_Fig5}
\end{subfigure}\hfill
\begin{subfigure}{0.3\linewidth}
\includegraphics[width=\linewidth]{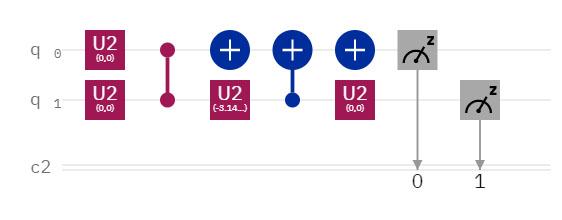} 
\caption{}
\label{qdctc_Fig6}
\end{subfigure}\hfill
\begin{subfigure}{0.3\linewidth}
\includegraphics[width=\linewidth]{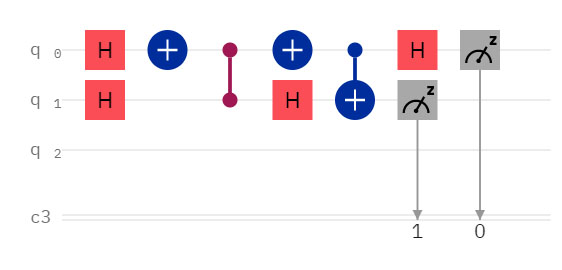} 
\caption{}
\label{qdctc_Fig6}
\end{subfigure}\hfill
\begin{subfigure}{0.3\linewidth}
\includegraphics[width=\linewidth]{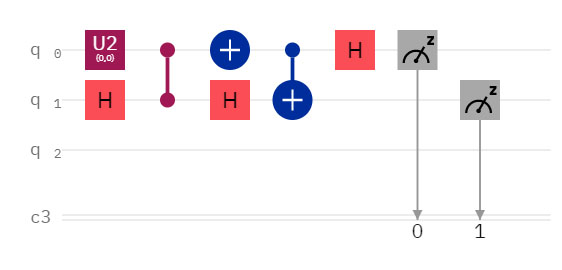} 
\caption{}
\label{qdctc_Fig6}
\end{subfigure}\hfill
\begin{subfigure}{0.3\linewidth}
\includegraphics[width=\linewidth]{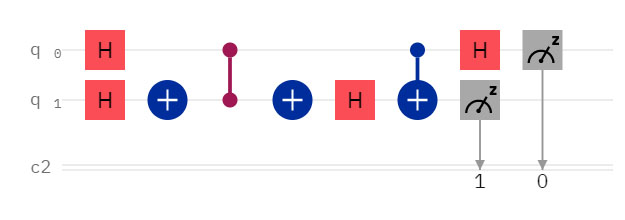} 
\caption{}
\label{qdctc_Fig6}
\end{subfigure}\hfill
\begin{subfigure}{0.3\linewidth}
\includegraphics[width=\linewidth]{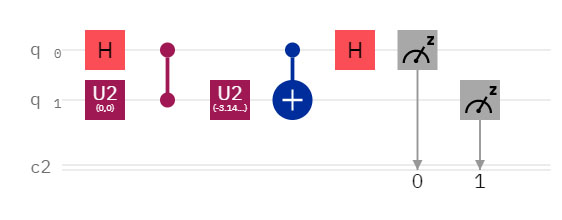} 
\caption{}
\label{qdctc_Fig6}
\end{subfigure}\hfill
\begin{subfigure}{0.3\linewidth}
\includegraphics[width=\linewidth]{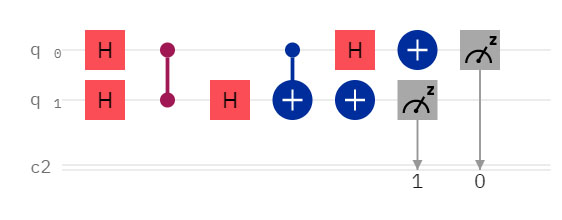} 
\caption{}
\label{qdctc_Fig6}
\end{subfigure}\hfill
\begin{subfigure}{0.3\linewidth}
\includegraphics[width=\linewidth]{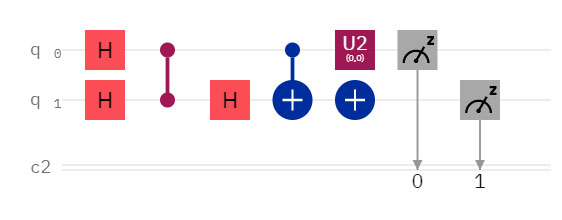} 
\caption{}
\label{qdctc_Fig6}
\end{subfigure}\hfill
\caption{ The original quantum circuits for Grover's search algorithm for the states of two qubit system \(\ket{00}\), \(\ket{01}\), \(\ket{10}\) and \(\ket{11}\) are shown in (a), (c), (e) and (g) respectively. The transpilled quantum circuits for \(\ket{00}\), \(\ket{01}\), \(\ket{10}\) and \(\ket{11}\) are shown in (b), (d), (f) and (h) respectively.}
\label{qdctc_Figp}
\end{figure*}

The two qubits system has \(2^2 = 4\) states, those are \(\ket{00}\), \(\ket{01}\), \(\ket{10}\), \(\ket{11}\). Here, we design the quantum circuits for OFPQS algorithm \cite{chuang1998experimental} for all of these four states and run those on IBMQ QASM simulator. The quantum circuits and the probability distributions for the states in each job are shown and explained. This signifies the fact that OFPQS algorithm \cite{chuang1998experimental}, helps us find the target states with highest probability [with respect to the other non-target states of that particular system (say, two qubits system here)]. The original  quantum circuits for  \(\ket{00}\), \(\ket{01}\), \(\ket{10}\), \(\ket{11}\) are (a), (c), (e), (g) and shown in [ Fig. \ref{qdctc_Figp}]. The transpilled quantum circuits for  \(\ket{00}\), \(\ket{01}\), \(\ket{10}\), \(\ket{11}\) are (b), (d), (f), (h) and shown in [Fig. \ref{qdctc_Figp}]. We run this quantum circuit on IBMQ QASM simulator. The probablities of the final states in this search algorithm for all the target states of the two qubit system are given in [ Table \ref{Table-I}].

\begin{table}
\caption{ This table shows the final outcomes and the probabilities of final outcome states after single iteration for the states of two qubit system }
\begin{tabular}{|c|c|c|}
\hline
Target state  &  Final outcome & Probabilty of final outcome\\
\hline
00 & 00 & 1.0\\
\hline
01 & 01 & 1.0\\
\hline
10 & 10 & 1.0\\
\hline
11 & 11 & 1.0\\
\hline
\end{tabular}
\label{Table-I}
\end{table}

\subsection{Three qubits system}
\begin{figure*}[!ht]
\centering
\begin{subfigure}{0.3\linewidth}
\includegraphics[width=\linewidth]{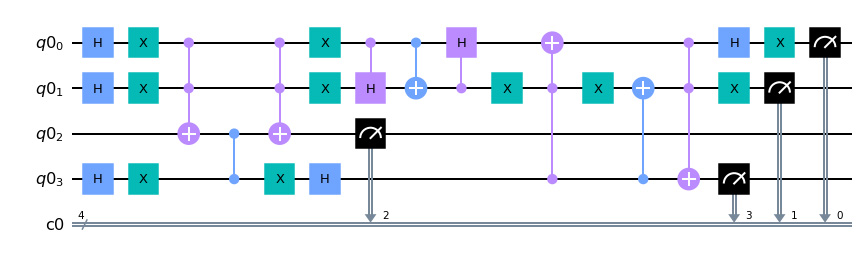} 
\caption{}
\label{qdctc_Fig5}
\end{subfigure}\hfill
\begin{subfigure}{0.3\linewidth}
\includegraphics[width=\linewidth]{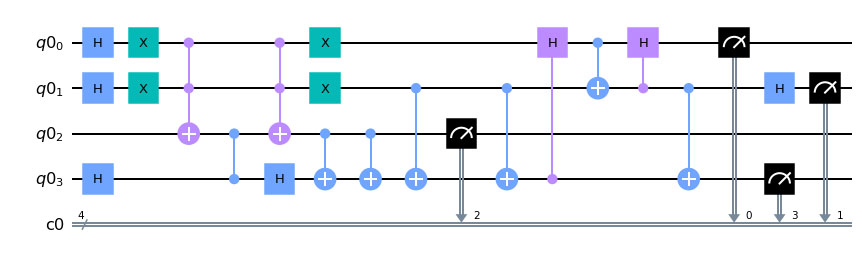} 
\caption{}
\label{qdctc_Fig6}
\end{subfigure}\hfill
\begin{subfigure}{0.3\linewidth}
\includegraphics[width=\linewidth]{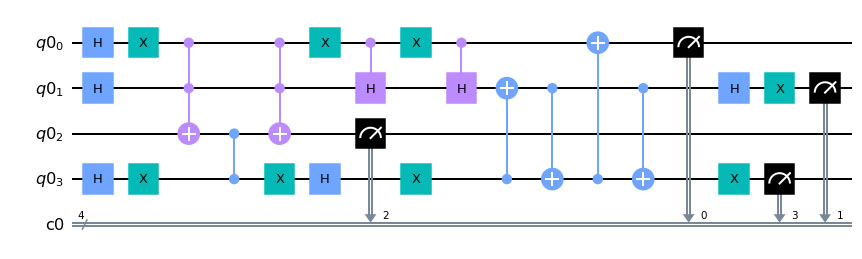} 
\caption{}
\label{qdctc_Fig6}
\end{subfigure}\hfill
\begin{subfigure}{0.3\linewidth}
\includegraphics[width=\linewidth]{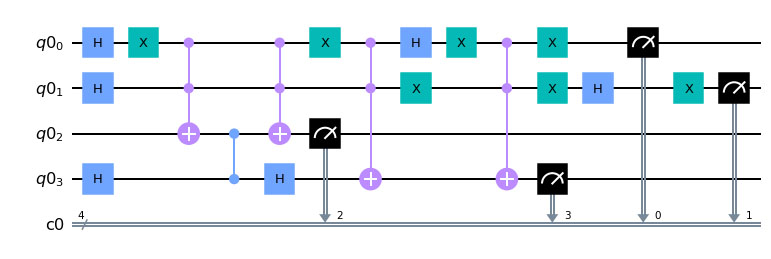}
\caption{}
\label{qdctc_Fig6}
\end{subfigure}\hfill
\begin{subfigure}{0.3\linewidth}
\includegraphics[width=\linewidth]{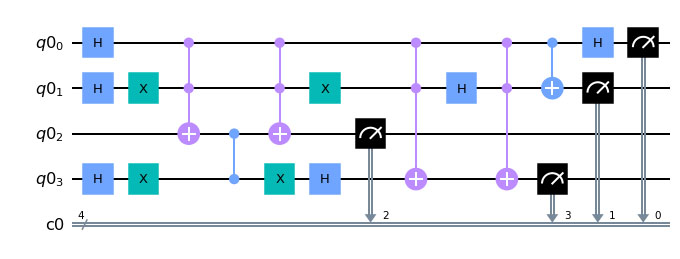} 
\caption{}
\label{qdctc_Fig6}
\end{subfigure}\hfill
\begin{subfigure}{0.3\linewidth}
\includegraphics[width=\linewidth]{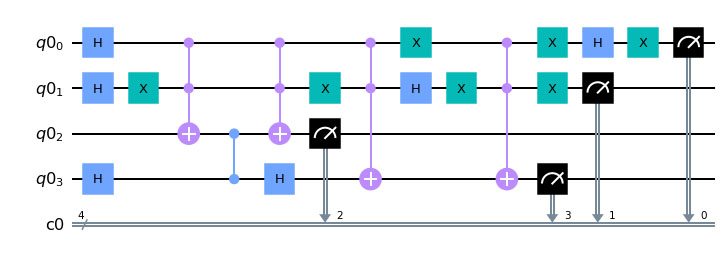} 
\caption{}
\label{qdctc_Fig6}
\end{subfigure}\hfill
\begin{subfigure}{0.3\linewidth}
\includegraphics[width=\linewidth]{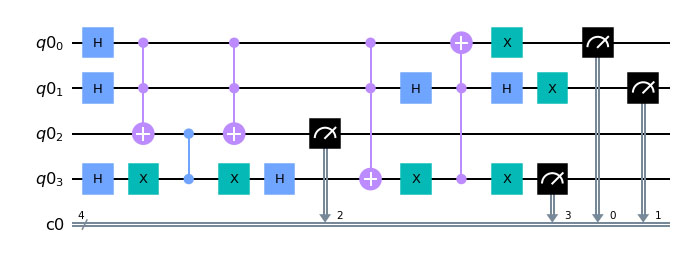} 
\caption{}
\label{qdctc_Fig6}
\end{subfigure}\hfill
\begin{subfigure}{0.3\linewidth}
\includegraphics[width=\linewidth]{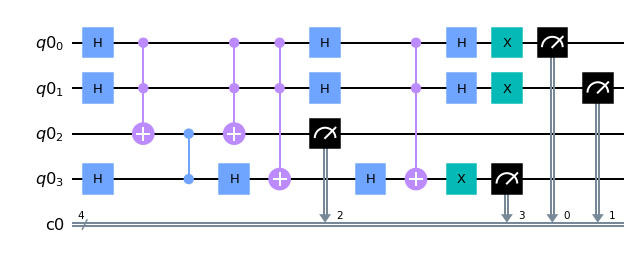} 
\caption{}
\label{qdctc_Fig6}
\end{subfigure}\hfill
\caption{{Quantum circuit for Grover's search algorithm for the states of three qubit system, (a) for \(\ket{000}\) state, (b) for \(\ket{001}\) state, (c) for \(\ket{010}\) state, (d) for \(\ket{011}\) state, (e) for \(\ket{100}\) state, (f) for \(\ket{101}\) state, (g) for \(\ket{110}\) state, (h) for \(\ket{111}\) state.}}
\label{qdctc_Figq}
\end{figure*}

The three qubits system has \(2^3 = 8\) states, those are \(\ket{000}\), \(\ket{111}\), \(\ket{110}\), \(\ket{101}\), \(\ket{100}\), \(\ket{011}\), \(\ket{101}\), \(\ket{010}\). Here, we design the quantum circuits for OFPQS algorithm for all of these eight states and run those on IBMQ QASM simulator. The quantum circuits and the probability distributions for the states in each job are shown and explained. This signifies the fact that OFPQS algorithm \cite{chuang1998experimental}, helps us find the target states with highest probability [with respect to the other non-target states of that particular system (say, three qubits system here)].The quantum circuits for all the states of three qubit systems are shown in Fig. \ref{qdctc_Figq}. We run this quantum circuit on IBMQ QASM simulator. The probablities of the final states in this search algorithm for all the target states of the three qubit system are given in Table \ref{Table-II}. The decomposition of multi-controlled qubit U( or Z) gate is shown in Fig. \ref{CIRCUIT new.jpg}. If there are $N_c$ number of control qubits and one target qubit, then one needs to insert $(N_c - 1)$ number of extra qubits in between the highest numbered control qubit and the target qubit, as depicted in Fig. \ref{CIRCUIT new.jpg}. In Fig. \ref{CIRCUIT new.jpg}, 'C'-designated qubits are controlled qubits, 'E'-designated qubits are extra qubits that are needed to be inserted in between the highest numbered control qubit $\&$ target qubit and 'q'-designated qubit is the target qubit.

\begin{table}
\caption{This table shows the final outcomes and the probabilities of final outcome states after single iteration for the states of three qubit system}
\begin{tabular}{|c|c|c|}
\hline
Target state  &  Final outcome & Probabilty of final outcome\\
\hline
000 & 000 & 1.0\\
\hline
001 & 001 & 0.987\\
\hline
010 & 010 & 0.943\\
\hline
011 & 011 & 0.983\\
\hline
100 & 100 & 0.971\\
\hline
101 & 101 & 1.0\\
\hline
110 & 110 & 0.955\\
\hline
111 & 111 & 0.988\\
\hline
\end{tabular}
\label{Table-II}
\end{table}

\subsection{Four qubits system}
\begin{figure*}[!ht]
\centering
\begin{subfigure}{0.3\linewidth}
\includegraphics[width=\linewidth]{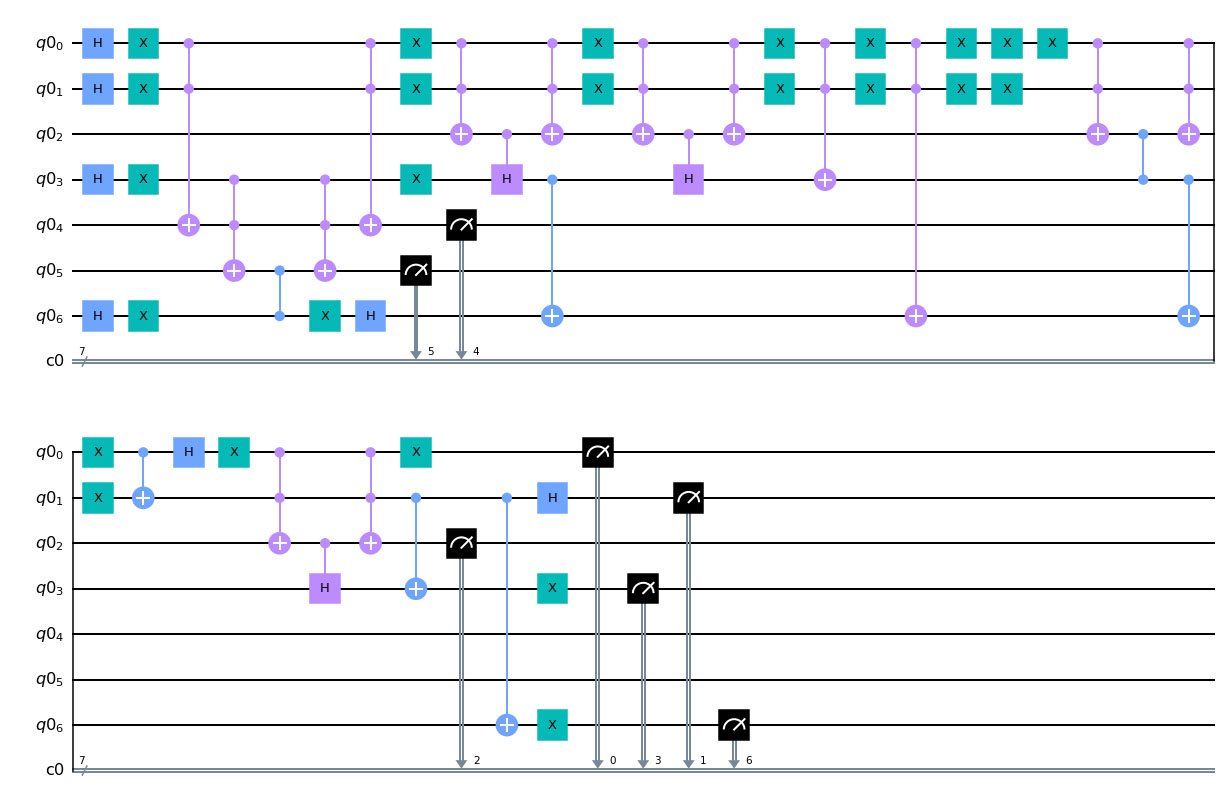} 
\caption{}
\label{qdctc_Fig5}
\end{subfigure}\hfill
\begin{subfigure}{0.3\linewidth}
\includegraphics[width=\linewidth]{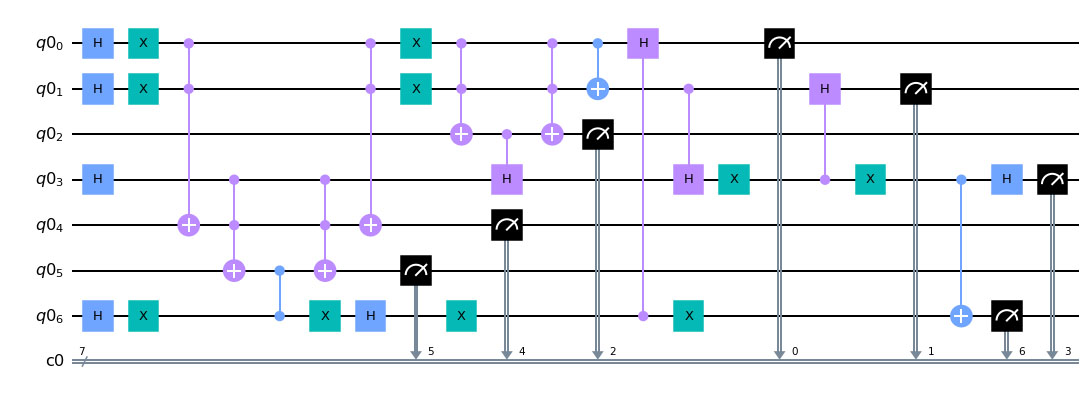} 
\caption{}
\label{qdctc_Fig6}
\end{subfigure}\hfill
\begin{subfigure}{0.3\linewidth}
\includegraphics[width=\linewidth]{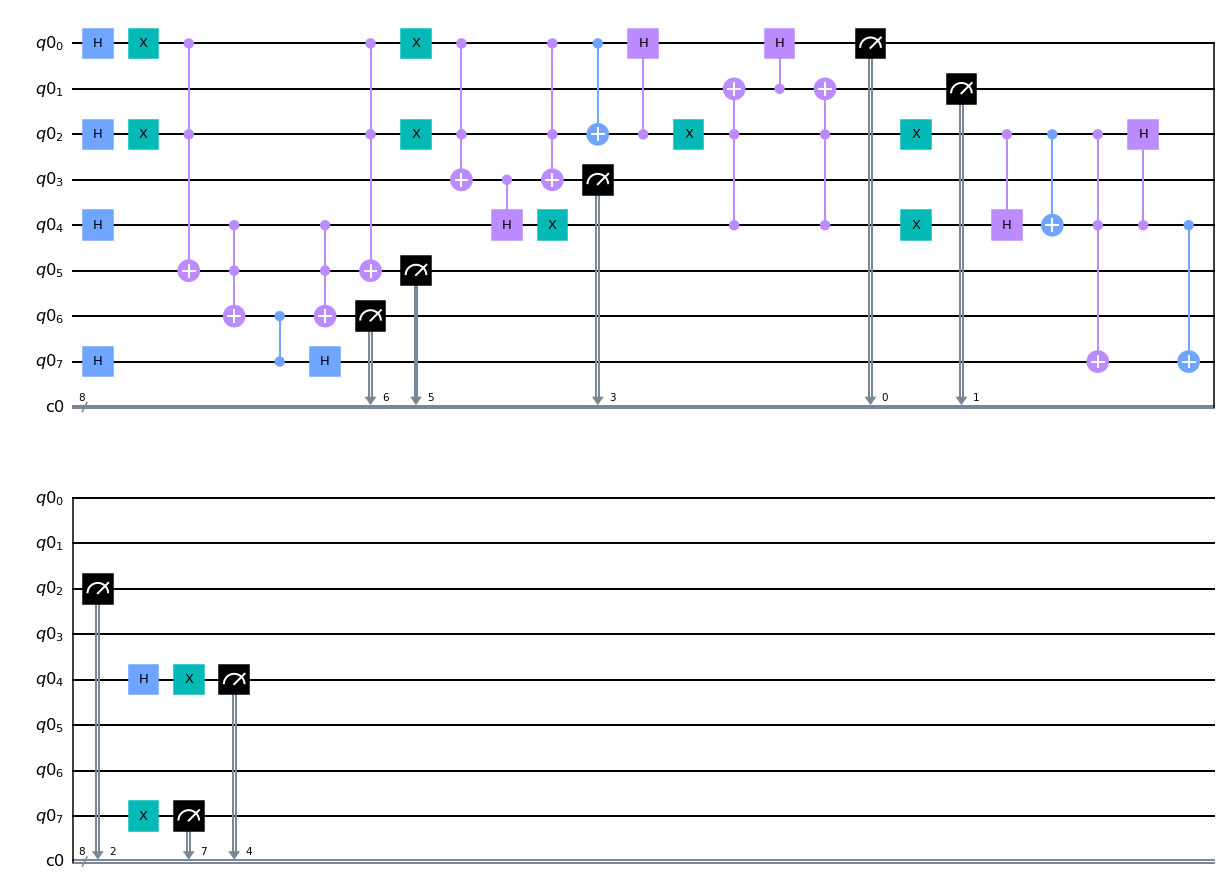} 
\caption{}
\label{qdctc_Fig6}
\end{subfigure}\hfill
\begin{subfigure}{0.3\linewidth}
\includegraphics[width=\linewidth]{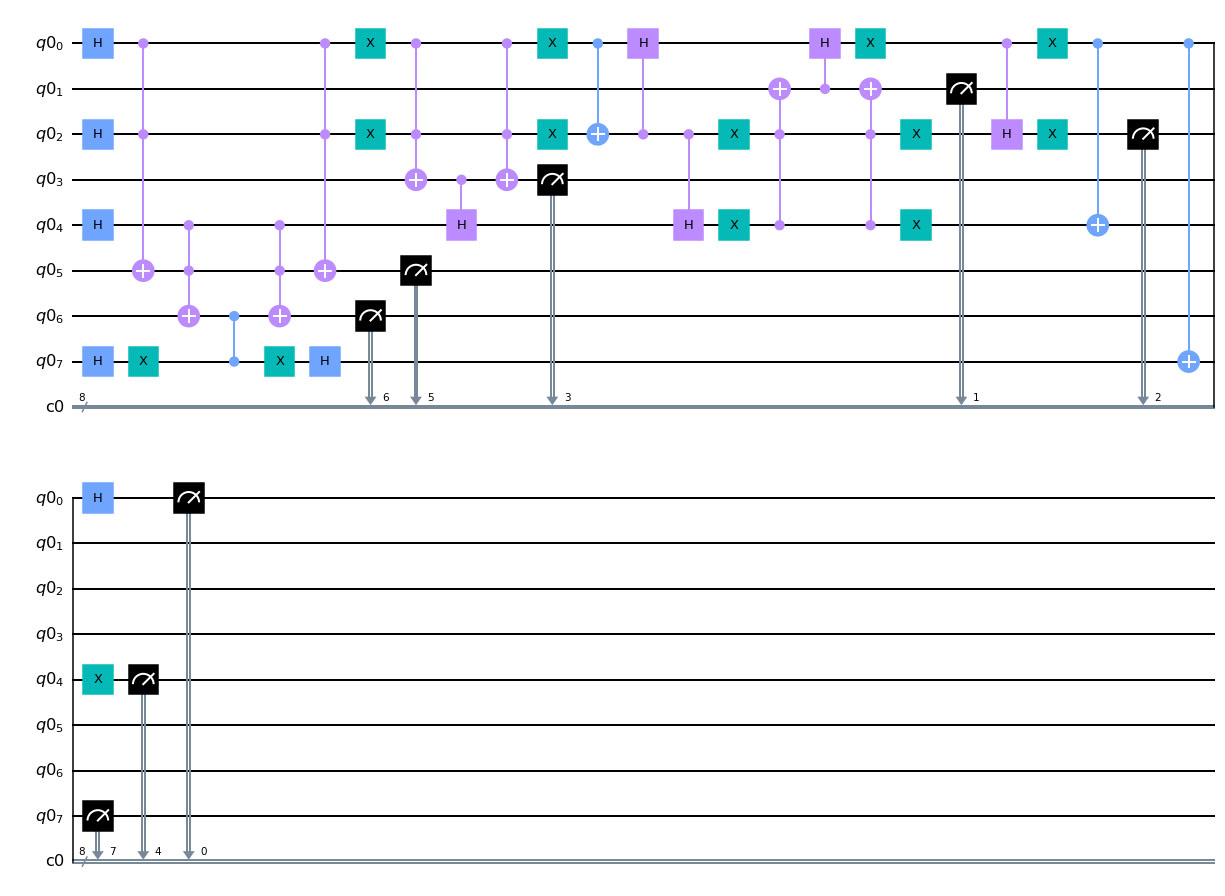}
\caption{}
\label{qdctc_Fig6}
\end{subfigure}\hfill
\begin{subfigure}{0.3\linewidth}
\includegraphics[width=\linewidth]{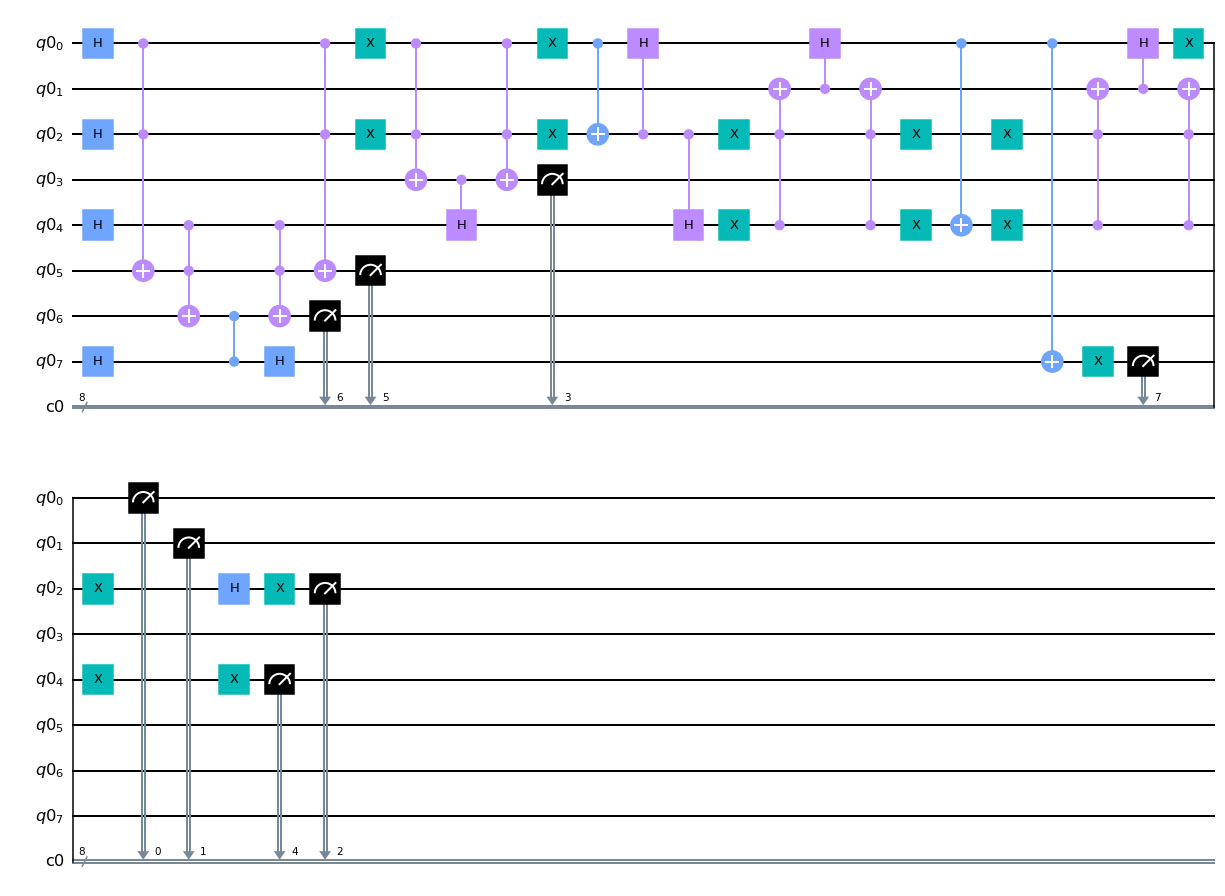} 
\caption{}
\label{qdctc_Fig6}
\end{subfigure}\hfill
\caption{{Quantum circuit for Grover's search algorithm for the states of four qubit system. (a) for \(\ket{0000}\) state. (b) for \(\ket{0010}\) state. (c) for \(\ket{0011}\) state. (d) for \(\ket{1110}\) state. (e) for \(\ket{1111}\) state.}}
\label{qdctc_Figm}
\end{figure*}

\begin{figure}
\centering
\includegraphics[width=\linewidth]{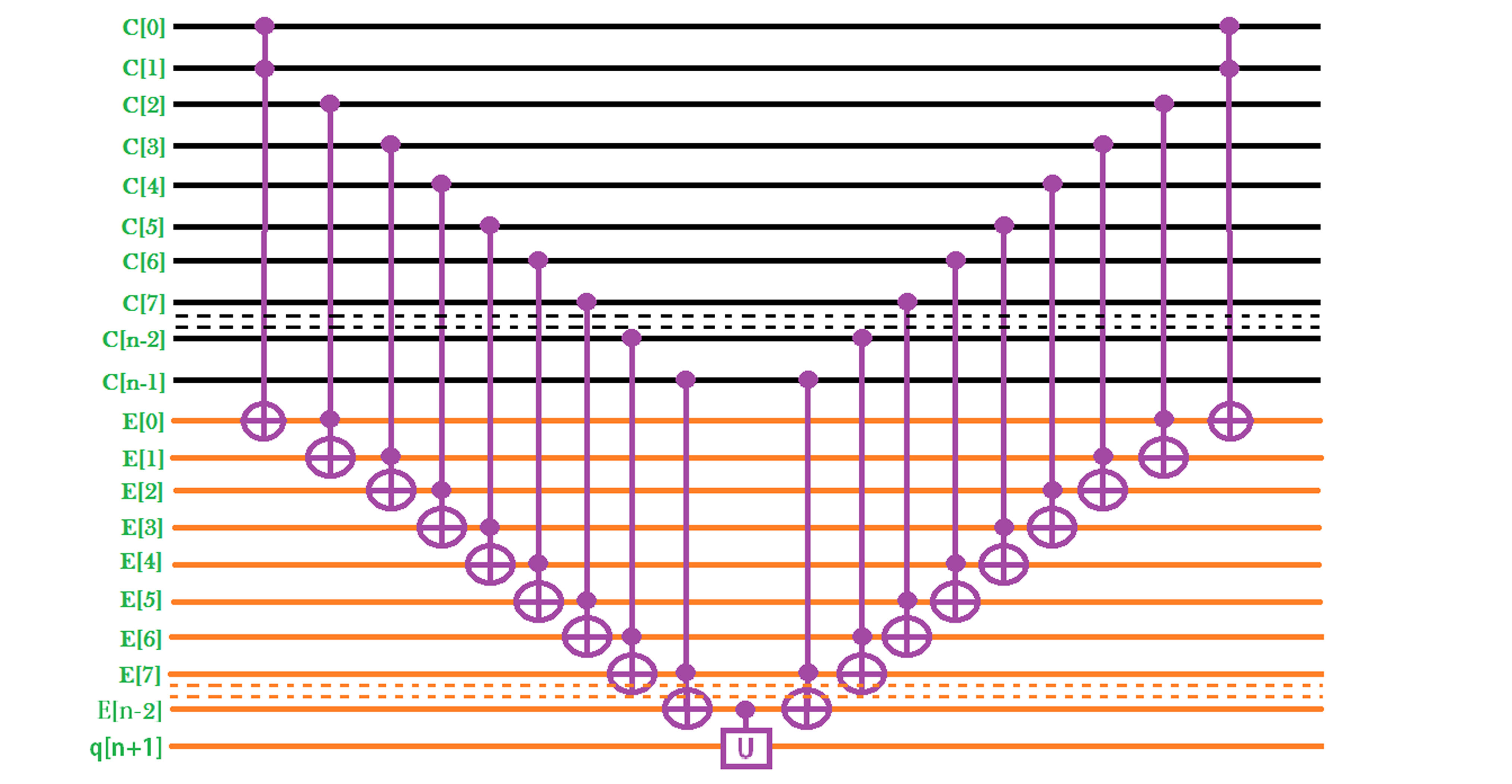}
\caption{\textbf{This figure depicts the decomposition of multi-controlled qubit U gate. Here, C[0], C[1],..., C[n-1] are controlled qubits and E[0], E[1],..., E[n-2] are extra qubits that are inserted in between the highest numbered controlled qubit C[n-1] $\&$ the target qubit q[n+1]. If the number of controlled qubits in a multi-controlled qubit U/Z gate is n, then the number of extra qubits that are needed to be inserted must be (n-1). }}
\label{CIRCUIT new.jpg}
\end{figure}

The four qubits system has \(2^4 = 16\) states, those are \textbf{\(\ket{0000}\)}, \textbf{\(\ket{0001}\)}, \textbf{\(\ket{0010}\)}, \textbf{\(\ket{0011}\)}, \textbf{\(\ket{1110}\)}, \textbf{\(\ket{1111}\)}, \textbf{\(\ket{1000}\)}, \textbf{\(\ket{1001}\)}, \textbf{\(\ket{0110}\)}, \textbf{\(\ket{0111}\)}, \textbf{\(\ket{1010}\)}, \textbf{\(\ket{1011}\)}, \textbf{\(\ket{0100}\)}, \textbf{\(\ket{0101}\)}, \textbf{\(\ket{1100}\)}, \textbf{\(\ket{1101}\)}. Here, we design the quantum circuits for OFPQS algorithm for all of these sixteen states and run those on IBMQ QASM simulator. The quantum circuits and the probability distributions for the states in each job are shown and explained. This signifies the fact that OFPQS algorithm \cite{chuang1998experimental}, helps us find the target states with highest probability [with respect to the other non-target states of that particular system (say, four qubits system here)]. The quantum circuits for (a) \(\ket{0000}\), (b) \(\ket{0010}\), (c) \(\ket{0011}\), (d) \(\ket{1110}\), (e)\(\ket{1111}\) are shown in [ Fig. \ref{qdctc_Figm}]. We run this quantum circuit on IBMQ QASM simulator. The probabilities of the final states in this search algorithm for the target states of the four qubit system are given in  [Table \ref{Table-III}]. The decomposition of multi-controlled qubit U( or Z) gate is shown in Fig. \ref{CIRCUIT new.jpg}.

\begin{figure*}[!ht]
\centering
\begin{subfigure}{0.3\linewidth}
\includegraphics[width=\linewidth]{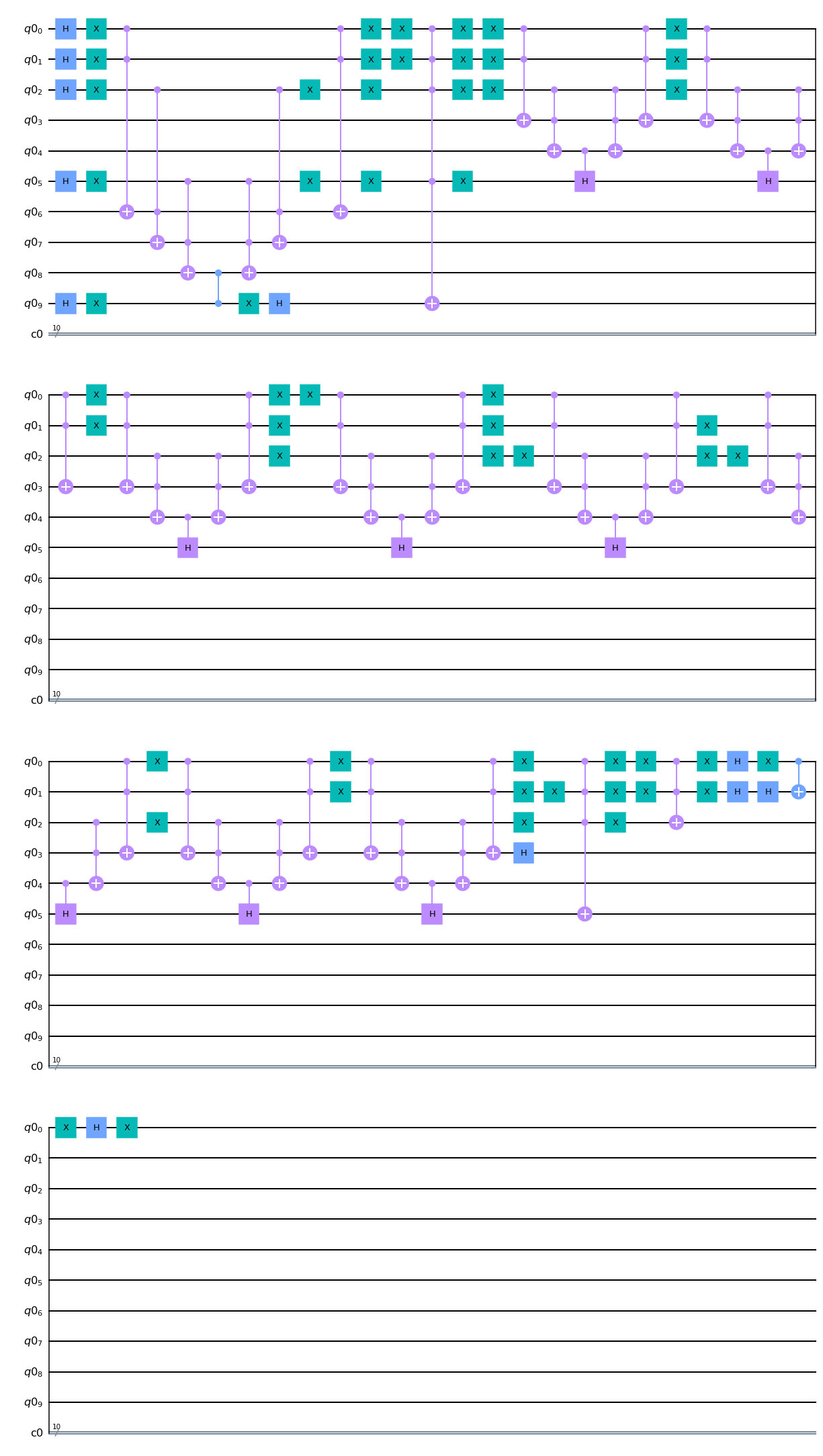} 
\caption{}
\label{qdctc_Fig5}
\end{subfigure}\hfill
\begin{subfigure}{0.3\linewidth}
\includegraphics[width=\linewidth]{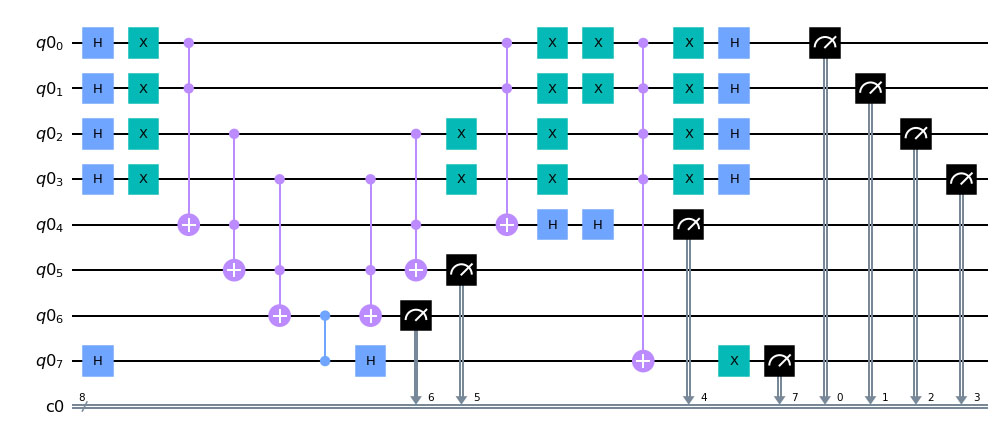} 
\caption{}
\label{qdctc_Fig6}
\end{subfigure}\hfill
\begin{subfigure}{0.3\linewidth}
\includegraphics[width=\linewidth]{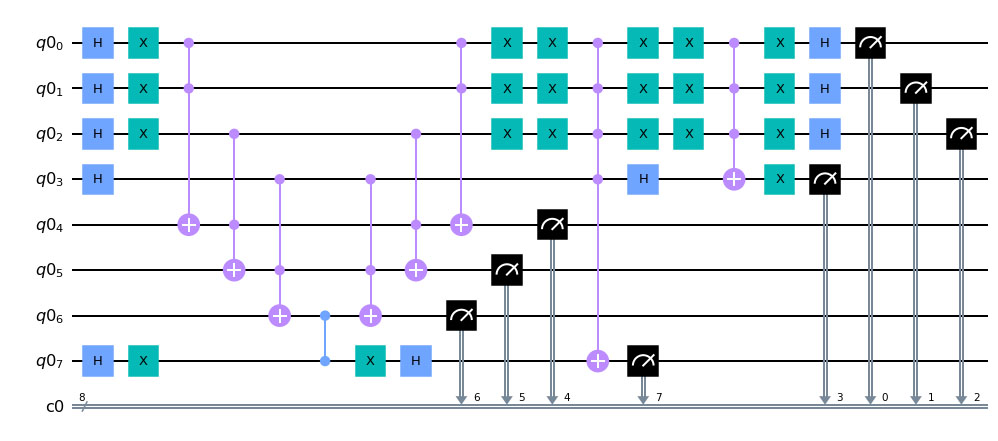} 
\caption{}
\label{qdctc_Fig6}
\end{subfigure}\hfill
\begin{subfigure}{0.3\linewidth}
\includegraphics[width=\linewidth]{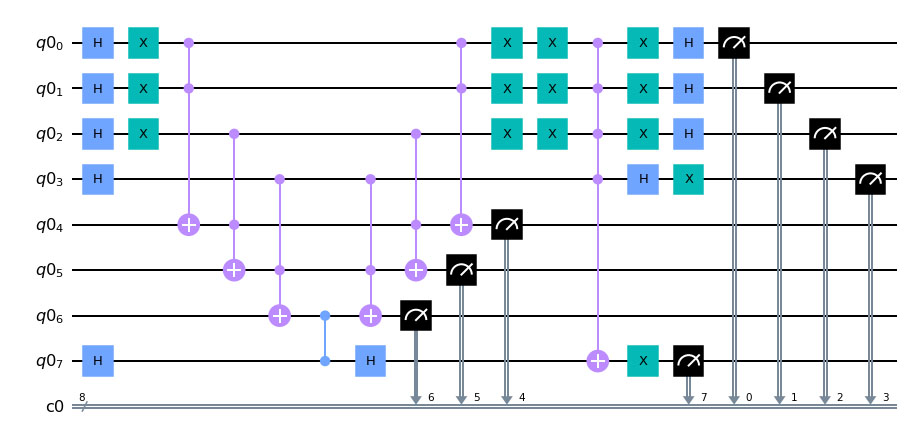}
\caption{}
\label{qdctc_Fig6}
\end{subfigure}\hfill
\begin{subfigure}{0.3\linewidth}
\includegraphics[width=\linewidth]{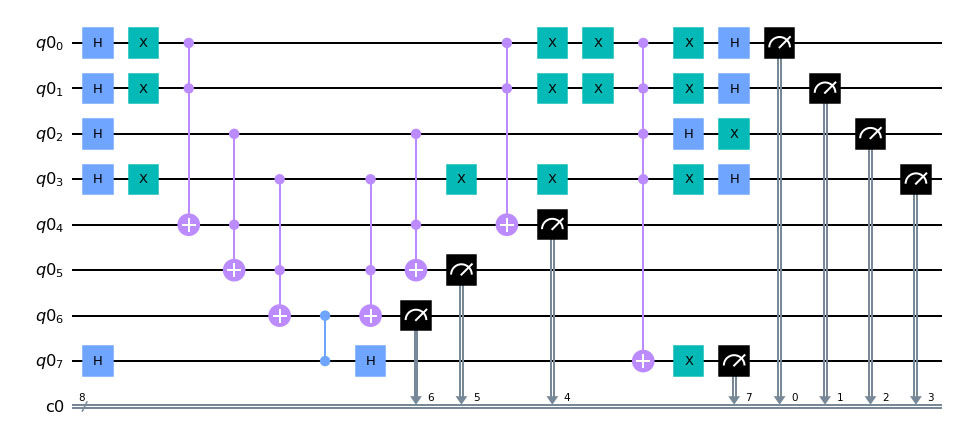} 
\caption{}
\label{qdctc_Fig6}
\end{subfigure}\hfill
\caption{ The Quantum circuits for Grover's search algorithm for the states of five qubit system \(\ket{00000}\), \(\ket{00001}\), \(\ket{00010}\), \(\ket{00011}\) , and \(\ket{00101}\) are (a), (b), (c), (d), and (e) respectively.}
\label{qdctc_Fign}
\end{figure*}

\begin{table}
\caption{ This table shows the final outcomes and the probabilities of final outcome states after single iteration for the states of four qubit system}
\begin{tabular}{|c|c|c|}
\hline
Target state  &  Final outcome & Probabilty of final outcome\\
\hline
0000 & 0000 & 1.000\\
\hline
0010 & 0010 & 0.991\\
\hline
0011 & 0011 & 0.974\\
\hline
1110 & 1110 & 0.979\\
\hline
1111 & 1111 & 0.985\\
\hline
\end{tabular}
\label{Table-III}
\end{table}

\begin{itemize}
\item The quantum circuit for target state \(\ket{1000}\) is depicted below. All four qubits are initially at \(\ket{0}\) state. The step-by-step operations and gates to reach the final state i.e., target state are depicted below. \(\ket{0000}\) [i.e., all four qubits are at \(\ket{0}\) initially] is the initial state. The quantum circuit for this marked state is 

$H_1\ H_2\ H_3\ H_4\ X_4\ C-AC-AC-Z_{1, 2, 3, 4}\ X_4\ H_4\ CCH_{1, 2, 3}\ AC-AC-H_{1, 4, 2}\ AC-AC-H_{1, 4, 3}\ C-AC-X_{1, 2, 3}\ AC-AC-H_{3, 2, 1}\ CX_{1, 2} AC-H_{2, 1}\ CX_{1, 2}\ CX_{1, 3}\ CX_{1, 4}\ H_1$.
Then we need to measure all the qubits. Here, C and AC stand for control and anti-control (anti-control is equivalent to XCX respectively. The subscripts 1,2,3,4 are used to mean the particular qubits like 1 for first qubit, 2 for second qubit etc. $H_{i}$ is meant by the application of H on i-th qubit), $AC-H_{i,j}$ is meant by the application of anti-control operation on i-th qubit and application of Hadamard operation on j-th qubit (Here, j-th qubit is the target qubit). We will use similar things throughout the paper.
\end{itemize}
\begin{itemize}
\item The quantum circuit for target state \(\ket{1001}\) is depicted below. All four qubits are initially at \(\ket{0}\) state. The step-by-step operations and gates to reach the final state i.e., target state are depicted below. \(\ket{0000}\) [ i.e., all four qubits are at \(\ket{0}\) initially] is the initial state. The quantum circuit for this marked state is $H_1\ H_2\ H_3\ H_4\ C-AC-AC-Z_{1, 2, 3, 4}\ H_4\ CCH_{1, 2, 3}\ AC-AC-H_{1, 2, 3}\ AC-CH_{1, 2, 3}\ AC-H_{2, 1}\ AC-H_{1, 2}\ CXX_{3, 4}\ CH_{1, 3}\ CX_{1,2}\ CX_{1, 4}\ H_1\ X_1\ X_4$. Then measure all the qubits.
\end{itemize}

\begin{itemize} 
\item\(\ket{0110}\): The quantum circuit for target state \(\ket{0110}\) is depicted below. All four qubits are initially at \(\ket{0}\) state. The step-by-step operations and gates to reach the final state i.e., target state are depicted below. \(\ket{0000}\) [i.e., all four qubits are at \(\ket{0}\) initially] is the initial state. The quantum circuit for this marked state is $H_1\ H_2\ H_3\ H_4\ X_4\ AC-C-C-Z_{1, 2, 3, 4}\ X_4\ H_4\ AC-AC-H_{1, 2, 3}\ CCH_{1, 2, 3}\ C-AC-H_{1, 2, 3}\ AC-H_{2, 1}\ CX_{1, 2}\ AC-H_{2, 1}\ AC-H_{3, 2}\ CX_{2, 3}\ CX_{2, 4}\  H_2\ X_3$. Then measure all the qubits.
\end{itemize}
\begin{itemize} 
\item The quantum circuit for target state \(\ket{0111}\) is depicted below. All four qubits are initially at \(\ket{0}\) state. The step-by-step operations and gates to reach the final state i.e., target state are depicted below. \(\ket{0000}\)[i.e., all four qubits are at \(\ket{0}\) initially] is the initial state. The quantum circuit for this marked state is $H_1\ H_2\ H_3\ H_4\ AC-C-C-Z_{1, 2, 3, 4}\  H_4\ AC-AC-H_{1, 2, 3}\ CCH_{1, 2, 3}\ C-AC-H_{1, 2, 3}\ AC-H_{2, 1}\ AC-AC-H_{1, 4, 2}\ CCX_{1, 2, 3}\ CCX_{1, 2, 4}\ CH_{2, 1}\ CX_{2, 3}\ CX_{2, 4}\ H_2\ \\
X_1\ X_2\ X_3$. Then measure all the qubits.
\end{itemize}
\begin{itemize}
\item The quantum circuit for target state \(\ket{1010}\) is depicted below. All four qubits are initially at \(\ket{0}\) state. The step-by-step operations and gates to reach the final state i.e., target state are depicted below. \(\ket{0000}\)[i.e., all four qubits are at \(\ket{0}\) initially] is the initial state. The quantum circuit for this marked state is $H_1\ H_2\ H_3\ H_4\ X_4\ C-AC-C-Z_{1, 2, 3, 4}\ X_4\ H_4\ AC-AC-H_{1, 2, 3}\ CCH_{1, 2, 3}\ AC-C-H_{1, 2, 3}\ AC-AC-H_{3, 2, 1}\ CCX_{1, 2, 3}\ CCX_{1, 2, 4}\ CH_{1, 2}\  CX_{1, 2}\ CX_{1, 3}\ CX_{1, 4}\ \\
AC-H_{2, 1}\ CX_{1, 2}\ H_1\ X_1\ X_2\ X_3$. Then measure all the qubits. 
\end{itemize}
\begin{itemize}
\item The quantum circuit for target state \(\ket{1011}\) is depicted below. All four qubits are initially at \(\ket{0}\) state. The step-by-step operations and gates to reach the final state i.e., target state are depicted below. \(\ket{0000}\) [i.e., all four qubits are at \(\ket{0}\) initially] is the initial state. The quantum circuit for this marked state is $H_1\ H_2\ H_3\ H_4\  C-AC-C-Z_{1, 2, 3, 4}\ H_4\ AC-AC-H_{1, 2, 3}\ CCH_{1, 2, 3}\ AC-C-H_{1, 2, 3}\ AC-AC-H_{3, 2, 1}\ AC-H_{1, 2}\ CCX_{1, 2, 3}\ CCX_{1, 2, 4}\ CH_{1, 2}\ CX_{1, 3}\ CX_{1, 4}\  H_1\ \\
X_1\ X_3\ X_4$. Then measure all the qubits.
\end{itemize}
\begin{itemize}
\item The quantum circuit for target state \(\ket{0100}\) is depicted below. All four qubits are initially at \(\ket{0}\) state. The step-by-step operations and gates to reach the final state i.e., target state are depicted below. \(\ket{0000}\)[i.e., all four qubits are at \(\ket{0}\) initially] is the initial state. The quantum circuit for this marked state is 

$H_1\ H_2\ H_3\ H_4\ X_4\ AC-C-AC-Z_{1, 2, 3, 4}\ X_4\ H_4\ AC-AC-H_{1, 2, 3}\ CCH_{1, 2, 3}\ C-AC-H_{1, 2, 3}\ AC-H_{2, 1}\ CX_{1, 2}\ AC-H_{2, 1}\ CX_{3, 4}\ CX_{3, 1}\ CX_{1, 2}\ CX_{1, 3}\ CX_{1, 4}\ AC-H_{2, 1}\ CX_{2, 4}\ H_2 $. Then measure all the qubits. 
\end{itemize}
\begin{itemize}
\item The quantum circuit for target state \(\ket{0101}\) is depicted below. All four qubits are initially at \(\ket{0}\) state. The step-by-step operations and gates to reach the final state i.e., target state are depicted below. \(\ket{0000}\)[i.e., all four qubits are at \(\ket{0}\) initially] is the initial state. The quantum circuit for this marked state is 
$H_1\ H_2\ H_3\ H_4\ AC-C-AC-Z_{1, 2, 3, 4}\ H_4\ AC-AC-H_{1, 2, 3}\ CCH_{1, 2, 3}\ C-AC-H_{1, 2, 3}\ AC-H_{2, 1}\ CX_{1, 2}\ AC-H_{2, 1}\ CX_{3, 1}\ CX_{1, 2}\ CX_{1, 3}\ AC-H_{2, 1}\ CX_{2, 4}\  H_2\ X_2\ X_4$. Then measure all the qubits.
\end{itemize}
\begin{itemize}
\item The quantum circuit for target state \(\ket{1100}\) is depicted below. All four qubits are initially at \(\ket{0}\) state. The step-by-step operations and gates to reach the final state i.e., target state are depicted below. \(\ket{0000}\)[i.e., all four qubits are at \(\ket{0}\) initially] is the initial state. The quantum circuit for this marked state is 

$H_1\ H_2\ H_3\ H_4\ X_4\ C-C-AC-Z_{1, 2, 3, 4}\ X_4\ H_4\ AC-AC-H_{1, 2, 3}\ C-AC-H_{1, 2, 3}\ AC-C-H_{1, 2, 3}\ AC-H_{3, 1}\ CX_{3, 1}\ AC-C-H_{1, 2, 3}\ AC-H_{3, 2}\ CX_{1, 2}\ CX_{1, 4}\  H_1\ X_2$. Then measure all the qubits. 
\end{itemize}
\begin{itemize}
\item The quantum circuit for target state \(\ket{1101}\) is depicted below. All four qubits are initially at \(\ket{0}\) state. The step-by-step operations and gates to reach the final state i.e., target state are depicted below. \(\ket{0000}\)[i.e., all four qubits are at \(\ket{0}\) initially] is the initial state. The quantum circuit for this marked state is 

$H_1\ H_2\ H_3\ H_4\ C-C-AC-Z_{1, 2, 3, 4}\ H_4\ AC-AC-H_{1, 2, 3}\ C-AC-H_{1, 2, 3}\ AC-C-H_{1, 2, 3}\ AC-H_{2, 1}\ AC-H_{1, 2}\ CX_{3, 1}\ CX_{3, 2}\ AC-H_{2, 3}\ CX_{1, 2}\ CX_{1, 4}\ H_1\ X_1\ X_2\ X_4$. Then measure all the qubits.
\end{itemize}
\subsection{Five qubits system}
The five qubits system has \(2^5 = 32\) states, those are \(\ket{00000}\), \(\ket{00001}\), \(\ket{00010}\), \(\ket{00011}\), \(\ket{00100}\), \(\ket{00101}\), \(\ket{01000}\), \(\ket{01001}\), \(\ket{10000}\), \(\ket{10001}\), \(\ket{11000}\), \(\ket{11001}\), \(\ket{10100}\), \(\ket{10101}\), \(\ket{10010}\), \(\ket{10011}\), \(\ket{01100}\), \(\ket{01101`}\), \(\ket{01010}\), \(\ket{01011}\), \(\ket{00110}\), \(\ket{00111}\), \(\ket{11100}\), \(\ket{11101}\), \(\ket{11010}\), \(\ket{11011}\), \(\ket{10110}\), \(\ket{10111}\), \(\ket{01110}\), \(\ket{01111}\), \(\ket{11110}\), \(\ket{11111}\). Here, we design the quantum circuits for OFPQS algorithm for some of these thirty-two states and run those on IBMQ QASM simulator. The quantum circuits and the probability distributions for the states in each job are shown and explained. This signifies the fact that OFPQS algorithm \cite{chuang1998experimental}, helps us find the target states with highest probability [ with respect to the other non-target states of that particular system (say, five qubits system here)].  The quantum circuit for (a) \(\ket{00000}\), (b) \(\ket{00001}\), (c) \(\ket{00010}\), (d) \(\ket{00011}\), (e) \(\ket{00101}\) are shown in [Fig. \ref{qdctc_Fign}]. We run this quantum circuit on IBMQ QASM simulator. The probabilities of the final states in this search algorithm for the target states of the five qubit system are given in [Table \ref{Table-IV}]. The decomposition of multi-controlled qubit U( or Z) gate is shown in Fig. \ref{CIRCUIT new.jpg}.

\begin{itemize}

\item The quantum circuit for target state \(\ket{01001}\) is depicted below. All five qubits are initially at \(\ket{0}\) state. The step-by-step operations and gates to reach the final state i.e., target state are depicted below. \(\ket{00000}\) [i.e., all five qubits are at \(\ket{0}\) initially] is the initial state. The quantum circuit for this marked state is $H_1\ H_2\ H_3\ H_4\ H_5\ AC-C-AC-AC-Z_{1, 2, 3, 4, 5}\ H_5\ AC-C-AC-AC-X_{1, 2, 3, 4, 5}\ H_4\ H_3\ H_1\ H_2\ X_2\ X_5$. Then measure all the qubits.

\end{itemize}

\begin{itemize}

\item The quantum circuit for target state \(\ket{10001}\) is depicted below. All five qubits are initially at \(\ket{0}\) state. The step-by-step operations and gates to reach the final state i.e., target state are depicted below. \(\ket{00000}\) [i.e., all five qubits are at \(\ket{0}\) initially] is the initial state. The quantum circuit for this marked state is $H_1\ H_2\ H_3\ H_4\ H_5\ C-AC-AC-AC-Z_{1, 2, 3, 4, 5}\ H_5\ C-AC-AC-AC-X_{1, 2, 3, 4, 5}\ H_4\ H_3\ H_1\ H_2\ X_1\ X_5$. Then measure all the qubits.

\end{itemize}

\begin{itemize}

\item The quantum circuit for target state \(\ket{11001}\) is depicted below. All five qubits are initially at \(\ket{0}\) state. The step-by-step operations and gates to reach the final state i.e., target state are depicted below. \(\ket{00000}\) [i.e., all five qubits are at \(\ket{0}\) initially] is the initial state. The quantum circuit for this marked state is $H_1\ H_2\ H_3\ H_4\ H_5\ C-C-AC-AC-Z_{1, 2, 3, 4, 5}\ H_5\ C-C-AC-AC-X\ H_4\ H_3\ H_1\ H_2\ X_1\ X_2\ X_5$. Then measure all the qubits.

\end{itemize}

\begin{itemize}

\item The quantum circuit for target state \(\ket{10101}\) is depicted below. All five qubits are initially at \(\ket{0}\) state. The step-by-step operations and gates to reach the final state i.e., target state are depicted below. \(\ket{00000}\) [i.e., all five qubits are at \(\ket{0}\) initially] is the initial state. The quantum circuit for this marked state is $H_1\ H_2\ H_3\ H_4\ H_5\ C-AC-C-AC-Z_{1, 2, 3, 4, 5}\ H_5\ C-AC-C-AC-X_{1, 2, 3, 4, 5}\ H_4\ H_3\ H_1\ H_2\ X_1\ X_3\ X_5$. Then measure all the qubits.

\end{itemize}

\begin{itemize}

\item The quantum circuit for target state \(\ket{10011}\) is depicted below. All five qubits are initially at \(\ket{0}\) state. The step-by-step operations and gates to reach the final state i.e., target state are depicted below. \(\ket{00000}\) [i.e., all five qubits are at \(\ket{0}\) initially] is the initial state. The quantum circuit for this marked state is $H_1\ H_2\ H_3\ H_4\ H_5\ C-AC-AC-C-Z_{1, 2, 3, 4, 5}\ H_5\ C-AC-AC-C-X_{1, 2, 3, 4, 5}\ H_4\ H_3\ H_1\ H_2\ X_1\ X_4\ X_5$. Then measure all the qubits.

\end{itemize}

\begin{itemize}

\item The quantum circuit for target state \(\ket{01101}\) is depicted below. All five qubits are initially at \(\ket{0}\) state. The step-by-step operations and gates to reach the final state i.e., target state are depicted below. \(\ket{00000}\) [i.e., all five qubits are at \(\ket{0}\) initially] is the initial state. The quantum circuit for this marked state is $H_1\ H_2\ H_3\ H_4\ H_5\ AC-C-C-AC-Z_{1, 2, 3, 4, 5}\ H_5\ AC-C-C-AC-X_{1, 2, 3, 4, 5}\ H_4\ H_3\ H_1\ H_2\ X_2\ X_3\ X_5$. Then measure all the qubits.

\end{itemize}

\begin{itemize}

\item The quantum circuit for target state \(\ket{01011}\) is depicted below. All five qubits are initially at \(\ket{0}\) state. The step-by-step operations and gates to reach the final state i.e., target state are depicted below. \(\ket{00000}\) [i.e., all five qubits are at \(\ket{0}\) initially] is the initial state. The quantum circuit for this marked state is $H_1\ H_2\ H_3\ H_4\ H_5\ AC-C-AC-C-Z_{1, 2, 3, 4, 5}\ H_5\ AC-C-AC-C-X_{1, 2, 3, 4, 5}\ H_4\ H_3\ H_1\ H_2\ X_2\ X_4\ X_5$. Then measure all the qubits.

\end{itemize}

\begin{itemize}

\item The quantum circuit for target state \(\ket{00111}\) is depicted below. All five qubits are initially at \(\ket{0}\) state. The step-by-step operations and gates to reach the final state i.e., target state are depicted below. \(\ket{00000}\) [i.e., all five qubits are at \(\ket{0}\) initially] is the initial state. The quantum circuit for this marked state is $H_1\ H_2\ H_3\ H_4\ H_5\ AC-AC-C-C-Z_{1, 2, 3, 4, 5}\ H_5\ AC-AC-C-C-X_{1, 2, 3, 4, 5}\ H_4\ H_3\ H_1\ H_2\ X_3\ X_4\ X_5$. Then measure all the qubits.

\end{itemize}

\begin{itemize}

\item The quantum circuit for target state \(\ket{11101}\) is depicted below. All five qubits are initially at \(\ket{0}\) state. The step-by-step operations and gates to reach the final state i.e., target state are depicted below. \(\ket{00000}\) [i.e., all five qubits are at \(\ket{0}\) initially] is the initial state. The quantum circuit for this marked state is $H_1\ H_2\ H_3\ H_4\ H_5\ C-C-C-AC-Z_{1, 2, 3, 4, 5}\ H_5\ C-C-C-AC-X_{1, 2, 3, 4, 5}\ H_4\ H_3\ H_1\ H_2\ X_1\ X_2\ X_3\ X_5$. Then measure all the qubits.

\end{itemize}

\begin{itemize}

\item The quantum circuit for target state \(\ket{11011}\) is depicted below. All five qubits are initially at \(\ket{0}\) state. The step-by-step operations and gates to reach the final state i.e., target state are depicted below. \(\ket{00000}\) [i.e., all five qubits are at \(\ket{0}\) initially] is the initial state. The quantum circuit for this marked state is $H_1\ H_2\ H_3\ H_4\ H_5\ C-C-AC-C-Z_{1, 2, 3, 4, 5}\ H_5\ C-C-AC-C-X_{1, 2, 3, 4, 5}\ H_4\ H_3\ H_1\ H_2\ X_1\ X_2\ X_4\ X_5$. Then measure all the qubits.

\end{itemize}

\begin{itemize}

\item The quantum circuit for target state \(\ket{10111}\) is depicted below. All five qubits are initially at \(\ket{0}\) state. The step-by-step operations and gates to reach the final state i.e., target state are depicted below. \(\ket{00000}\) [i.e., all five qubits are at \(\ket{0}\) initially] is the initial state. The quantum circuit for this marked state is $H_1\ H_2\ H_3\ H_4\ H_5\ C-AC-C-C-Z_{1, 2, 3, 4, 5}\ H_5\ C-AC-C-C-X_{1, 2, 3, 4, 5}\ H_4\ H_3\ H_1\ H_2\ X_1\ X_3\ X_4\ X_5$. Then measure all the qubits.

\end{itemize}

\begin{itemize}

\item The quantum circuit for target state \(\ket{01111}\) is depicted below. All five qubits are initially at \(\ket{0}\) state. The step-by-step operations and gates to reach the final state i.e., target state are depicted below. \(\ket{00000}\) [i.e., all five qubits are at \(\ket{0}\) initially] is the initial state. The quantum circuit for this marked state is $H_1\ H_2\ H_3\ H_4\ H_5\ AC-C-C-C-Z_{1, 2, 3, 4, 5}\ H_5\ AC-C-C-C-X_{1, 2, 3, 4, 5}\ H_4\ H_3\ H_1\ H_2\ X_2\ X_3\ X_4\ X_5$. Then measure all the qubits.

\end{itemize}

\begin{itemize}

\item The quantum circuit for target state \(\ket{11111}\) is depicted below. All five qubits are initially at \(\ket{0}\) state. The step-by-step operations and gates to reach the final state i.e., target state are depicted below. \(\ket{00000}\) [i.e., all five qubits are at \(\ket{0}\) initially] is the initial state. The quantum circuit for this marked state is $H_1\ H_2\ H_3\ H_4\ H_5\ C-C-C-C-Z_{1, 2, 3, 4, 5}\ H_5\ C-C-C-C-X_{1, 2, 3, 4, 5}\ H_4\ H_3\ H_1\ H_2\ X_1\ X_2\ X_3\ X_4\ X_5$. Then measure all the qubits.

\end{itemize}

\begin{table}
\caption{This table shows the final outcomes and the probabilities of final outcome states after single iteration for the states of five qubit system}
\begin{tabular}{|c|c|c|}
\hline
Target state  &  Final outcome & Probabilty of final outcome\\
\hline
00000 & 00000 & 1.000\\
\hline
00001 & 00001 & 0.996\\
\hline
00010 & 00010 & 0.988\\
\hline
00011 & 00011 & 0.995\\
\hline
00101 & 00101 & 0.989\\
\hline
\end{tabular}
\label{Table-IV}
\end{table}

\section{Conclusion}
To conclude, we have experimentally demonstrated here the optimal fixed-point quantum search algorithm using the IBMQ QASM simulator. We design the equivalent quantum circuits (of the algorithm) for various target states of two, three, four and five qubit systems and run the quantum circuits on IBMQ QASM simulator. We explicate the working of quantum algorithm by fixing a target quantum state, which has been prepared in our experiment. We discuss some comments about the results obtained in all these cases. The two or more iteration cases involve a large number of gates which introduce more noise and decoherence in the system.

\section{Acknowledgement}
B.D. and K.G. are grateful to IISER Kolkata for gracious hospitality. B.K.B. is financially supported by Institute Fellowship. The authors are extremely grateful to IBM team and IBM QE project. The discussions and opinions developed in this paper are only those of the authors and do not reflect the opinions of IBM or IBM QE team.







\end{document}